\documentclass[aps,prd,twocolumn,preprintnumbers,superscriptaddress,nofootinbib,floatfix]{revtex4-1}

\usepackage{amsmath}
\usepackage{bm}
\usepackage{amsfonts}
\usepackage{graphicx}
\usepackage{epstopdf}
\usepackage{hyperref}
\usepackage{array}
\usepackage[utf8]{inputenc}

\usepackage{soul}
\usepackage{color}

\enlargethispage{\baselineskip}

\hypersetup{
     colorlinks   = true,
     citecolor    = blue
}

\everymath{\displaystyle}

\newcommand{\MP}{M_{\rm PL}}

\newcommand{\HH}{\mathcal{H}}

\renewcommand\({\left(}
\renewcommand\){\right)}
\renewcommand\[{\left[}
\renewcommand\]{\right]}
\newcommand{\be}{\begin{equation}}
\newcommand{\ee}{\end{equation}}
\newcommand{\bea}{\begin{eqnarray}}
\newcommand{\eea}{\end{eqnarray}}

\begin{document}

\title{\bf  Late time transitions in the quintessence field and the $H_0$ tension}

\newcommand{\FIRSTAFF}{\affiliation{Department of Physics and Astronomy, Uppsala University, L\"agerhyddsv\"agen 1, 75120 Uppsala, Sweden}}
\newcommand{\SECONDAFF}{\affiliation{Jodrell Bank Center for Astrophysics, School of Physics and Astronomy,
University of Manchester, Oxford Road, Manchester, M13 9PL, UK}}
\newcommand{\THIRDAFF}{\affiliation{Nordita, KTH Royal Institute of Technology and Stockholm University, Roslagstullsbacken 23, 10691 Stockholm, Sweden}}
\newcommand{\FOURTHAFF}{\affiliation{Gravitation Astroparticle Physics Amsterdam (GRAPPA), Institute for Theoretical Physics Amsterdam and Delta Institute for Theoretical Physics, University of Amsterdam, Science Park 904, 1098 XH Amsterdam, The Netherlands}}

\author{Eleonora Di Valentino}\email[Electronic address: ]{eleonora.divalentino@manchester.ac.uk}\SECONDAFF
\author{Ricardo Z. Ferreira}\email[Electronic address: ]{ricardo.zambujal@su.se} \THIRDAFF
\author{Luca Visinelli}\email[Electronic address: ]{luca.visinelli@physics.uu.se} \FIRSTAFF \THIRDAFF \FOURTHAFF
\author{Ulf Danielsson}\email[Electronic address: ]{ulf.danielsson@physics.uu.se}\FIRSTAFF

\preprint{UUITP-23/19}
\date{\today}

\begin{abstract}
We consider a quintessence field which transitions from a matter-like to a cosmological constant behavior between recombination and the present time. We aim at easing the tension in the measurement of the present Hubble rate, and we assess the $\Lambda$CDM model properly enlarged to include our quintessence field against cosmological observations. The model does not address the scope we proposed. This result allows us to exclude a class of quintessential models as a solution to the tension in the Hubble constant measurements.
\end{abstract}

\maketitle

\section{Introduction}
\label{intro}

The nature of dark matter and dark energy remains a completely open question. While we do not lack theoretically well-motivated dark matter models, this is not the case for dark energy. The most minimal solution, a cosmological constant, remains the most appealing possibility although its very small size is difficult to understand from a purely theoretical perspective. In the context of a fundamental theory of quantum gravity, such as string theory, there are also deep conceptual problems related with the cosmological horizon in an accelerating cosmology. In particular, one of the main issues with string theory is its preference for Anti de Sitter (AdS) vacua, which is in sharp contrast with the observational evidence for a positive cosmological constant. Although some solutions with long lived metastable vacua have been proposed~\cite{Kachru:2003aw}, this has recently raised the problem that it is difficult to write self-consistent models of quantum gravity that live in the landscape of the string theory instead of the swampland~\cite{Danielsson:2018ztv, Obied:2018sgi, Kinney:2018nny}.

A way to circumvent such problem is to consider a time-varying dark energy modelled by a slowly rolling quintessence field. However, the simplest quintessence scenarios, a massive scalar field $\phi$, requires the field to be extremely light, $m_\phi \lesssim 10^{-33}\,$eV, thus raising questions about the stability of such a potential. One possibility that moves in this direction relies on an ``ultra-light'' axion arising within the context of string theory, in the so-called axiverse~\cite{Svrcek:2006yi, Arvanitaki:2009fg, Cicoli:2012sz, Marsh:2013taa, Visinelli:2018utg}. In Ref.~\cite{Banerjee:2018qey} yet another possibility was considered in which the time-dependence of the dark energy component is granted by the expansion in an extra dimension.

In this paper we investigate whether a quintessence field with a rapidly varying transition in the equation of state might alleviate the $4.4\sigma$ tension between local measurements of the Hubble constant \cite{Riess:2019cxk} and the value obtained using the CMB~\cite{Abbott:2018wog} in a $\Lambda$CDM scenario. This turns out to be surprisingly difficult. We consider a scalar field $\phi$ whose equation of state tracks the evolution of cold matter around recombination, transitioning to a generic equation of state $w_{\phi0}$ at a later time, thus providing a late time ``boost'' to the expansion, as allowed by the local measurements of $H_0$ and BAO data~\cite{Bernal:2018cxc}. We find that there is hardly any effect on the discrepancy between the different measurements of the Hubble constant. Our results suggest, in line with earlier work on the subject~\cite{Poulin:2018cxd, DEramo:2018vss, Yang:2018euj, Yang:2018uae, Yang:2018qmz, Alexander:2019rsc, Agrawal:2019dlm}, that the resolution to the problem needs to be found in the early Universe, possibly through a modified sound horizon at recombination (see {\it e.g.} Refs.~\cite{Evslin:2017qdn, Aylor:2018drw, Agrawal:2019lmo}).

One example of an explicit realisation of this dark matter to dark energy transition is to consider that the massive field $\phi$ contains metastable minima at the bottom of its potential. These minima could naturally come from higher harmonic corrections, other instanton contributions or simply thermal effects~\cite{Kaloper:2008fb, Silverstein:2008sg, Jaeckel:2016qjp, DAmico:2018mnx, Kobayashi:2018nzh,Baratella:2018pxi}. The field initially oscillates in its potential, behaving as a matter fluid with equation of state $w_\phi \simeq 0$. When the amplitude of the oscillations becomes comparable to the size of the barriers, the field gets trapped in one of those minima and starts acting as a dark energy fluid with $w_\phi \simeq -1$, until it eventually jumps/tunnels to the next minima. This is an entertaining possibility with peculiar predictions such as bubble formation or enhanced dark energy perturbations, which might be detectable in future dark energy surveys~\cite{Abbott:2018wog}.

This work is organised as follows. In Sec.~\ref{sec:Phenomodel} we provide the details for the phenomenological model that is later considered in the data analysis, and we describe an explicit realization of such a phenomenological description in a particle physics model. In Sec.~\ref{sec:backgroundperturbations} we provide the relevant equations for the background and perturbations used in the numerical analysis, which is described in Sec.~\ref{sec:Method} along with the datasets used and the parameter space explored. Results are presented in Sec.~\ref{sec:Results}. We conclude with the final remarks in Sec.~\ref{sec:Conclusions}.

\section{The model \label{sec:Phenomodel}}
 
\subsection{Modelling the quintessence phenomenology \label{sec:pheno}} 

We consider a scenario in which the equation of state of a quintessential field transitions from being matter-like, tracking the evolution of cold dark matter, to that of a fluid with equation of state $w_{\phi 0}$ at later times. The simplest parametrisation of such a behaviour is depicted by the following evolution of the energy density in the quintessence field
\be
	\rho_\phi(a) \!=\! \rho_{\phi,0}\(\frac{a}{a_*}\)^{\!-3}\!\!\[\Theta\(a - a_*\) \(\frac{a}{a_*}\)^{-3w_{\phi0}}\!\!\! \!+\! \Theta\(a_* \!-\! a\)\]\,,
	\label{eq:abrupt_rhophi}
\ee
where $\Theta(x)$ is the Heaviside step function on the variable $x$, $\rho_{\phi,0}$ is the value of $\rho_\phi$ at present time and we assumed an instantaneous freezing of the oscillations of the field $\phi$ at the time of the transition, where the scale factor is $a_*$.

\begin{figure}
	\centering
	\includegraphics[width=8cm]{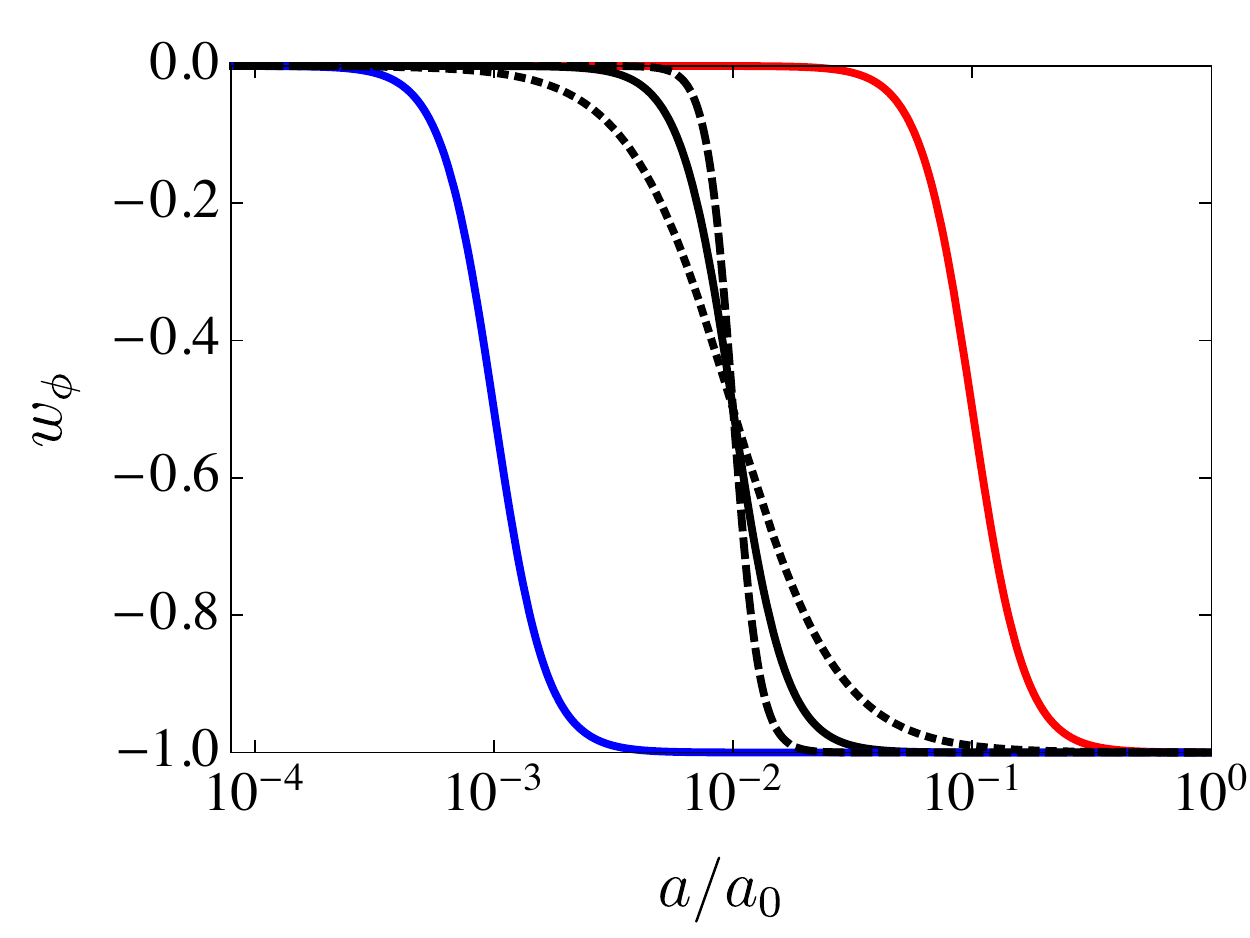}
	\caption{The equation of state $w_{\rm \phi \,eff}(a)$ in Eq.~\eqref{eq:weff} as a function of the scale factor $a$, for different values of the parameters $a_*$ and $\Delta$. See the text for additional details.}
	\label{fig:wphi}
\end{figure}

Instead of the parametrisation in Eq.~\eqref{eq:abrupt_rhophi}, we model a smoother transition by considering the effective equation of state
\be
    w_{\rm \phi \,eff}(a) = \frac{w_{\phi 0}}{1 + \(\frac{a}{a_*}\)^{-\frac{2}{\Delta}}}\,,
    \label{eq:weff}
\ee
where the scale factor $a_*$ controls the time of the transition and the constant $\Delta$ defines its duration, so that a shorter $\Delta$ corresponds to a shorter transition period. In Fig.~\ref{fig:wphi} we show the evolution of the equation of state in Eq.~\eqref{eq:weff} for $w_{\phi 0}=-1$ and for different choices of the parameters $a_* = 10^{-1}$ (red), $a_* = 10^{-2}$ (black), and $a_* = 10^{-3}$ (blue). The smaller the value of $a_*$, the earlier the transition occurs. For $a_* = 10^{-2}$ we also show the evolution for different choices of the width $\Delta = 1$ (dotted black), $\Delta = 0.5$ (solid black), and $\Delta = 0.25$ (dashed black). A smaller value of $\Delta$ corresponds to a sharper transition.

\begin{figure}
	\centering
	\includegraphics[width=8cm]{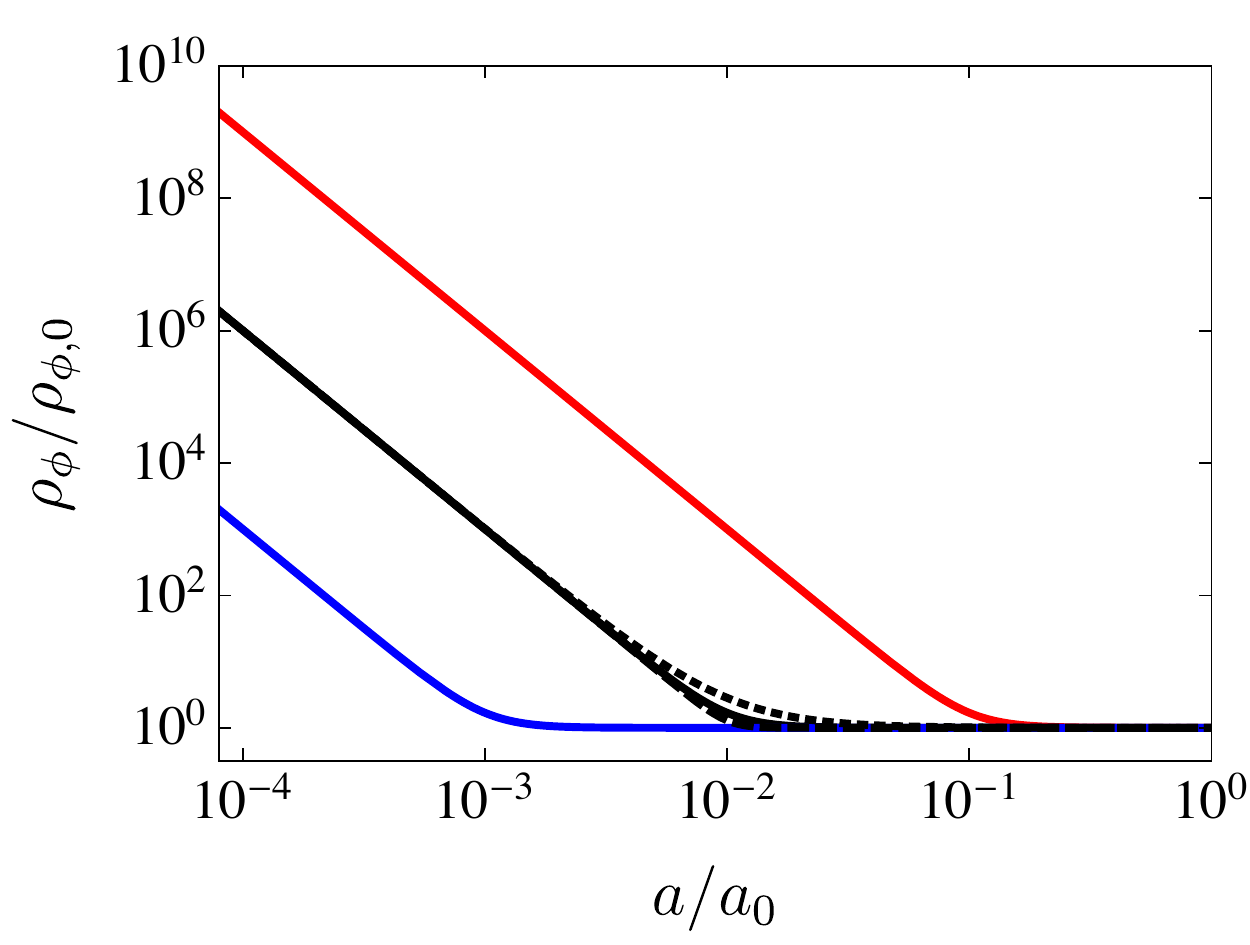}
	\caption{The energy density of the quintessence field $\rho_{\rm \phi \,eff}(a)$ in Eq.~\eqref{eq:energydensity} as a function of the scale factor $a$, for the same values of the parameters $a_*$ and $\Delta$ as in Fig.~\ref{fig:wphi}. See the text for additional details.}
	\label{fig:rhophi}
\end{figure}

Integrating the non-interacting continuity equation for $\rho_\phi$ with the effective equation of state in Eq.~\eqref{eq:weff} gives
\be
	\rho_\phi = \rho_{\phi, 0}\, \(\frac{a}{a_0}\)^{-3(1+w_{\phi 0})}\,\[\frac{1 + \(\frac{a_0}{a_*}\)^{-\frac{2}{\Delta}}}{1 + \(\frac{a}{a_*}\)^{-\frac{2}{\Delta}}}\]^{\frac{3\,\Delta\,w_{\phi 0}}{2}}\,,
	\label{eq:energydensity}
\ee
where $a_0$ and $\rho_{\phi, 0}$ are the present value of the scale factor and of the energy density in quintessence, respectively. In the following we fix $a_0 = 1$. The expression in Eq.~\eqref{eq:energydensity} shows the correct behaviour $\rho_\phi \propto a^{-3}$ for $a \ll a_*$ and $a^{-3(1+w_{\phi 0})}$ for $a \gg a_*$. In Fig.~\ref{fig:rhophi} we show the evolution of the energy density $\rho_\phi / \rho_{\phi, 0}$ as a function of the scale factor $a$, for the same parameters used in Fig.~\ref{fig:wphi}. 
In this work, we assess the validity of the quintessence model described by the equation of state in Eq.~\eqref{eq:weff} against various datasets, as we discuss in depth in Sec.~\ref{sec:Method} below. The same set of equations has been used for other works in which the interaction between the dark matter and the dark energy is constrained~\cite{Costa:2013sva}, as well as in models of ``Early Dark Energy''~\cite{Poulin:2018cxd}, dark energy with a phantom-like equation of state~\cite{DiValentino:2016hlg, DiValentino:2017zyq}, interacting dark energy~\cite{Kumar:2016zpg, DiValentino:2017iww}, or vacuum phase transitions~\cite{DiValentino:2017rcr, Khosravi:2017hfi}. The model we present here is complementary to the Early Dark Energy model of Ref.~\cite{Poulin:2018cxd}, in which a quintessence field that behaves as a cosmological constant at early times transitions to a matter-like or radiation-like behaviour today.

\subsection{An explicit realisation of the model  \label{sec:realization}}

There are several ways in which the field can get trapped in a metastable vacua. Here we discuss one possible implementation of the model where we add an extra scalar field $\phi$ of mass $m$ on top of the content of the $\Lambda$CDM model. The scalar field Lagrangian is given by
\begin{eqnarray}
    {\cal L} &=& -\frac 1 2 \partial_\mu \phi \partial^\mu \phi - V(\phi) \label{Lag} \, ,
    \label{eq:action}\\
    V(\phi) &=& \frac 1 2 m^2 \phi^2 + V_{\rm osc}(\phi)\,,
\end{eqnarray}
thus enforcing the scalar field to satisfy the Klein-Gordon equation
\be
    \ddot \phi +3H\dot\phi - \nabla^2\phi + m^2\phi + \frac{\partial V_{\rm osc}}{\partial \phi} = 0 \, .
    \label{eq:KGequation}
\ee

\begin{figure}
	\centering
	\includegraphics[width=8cm]{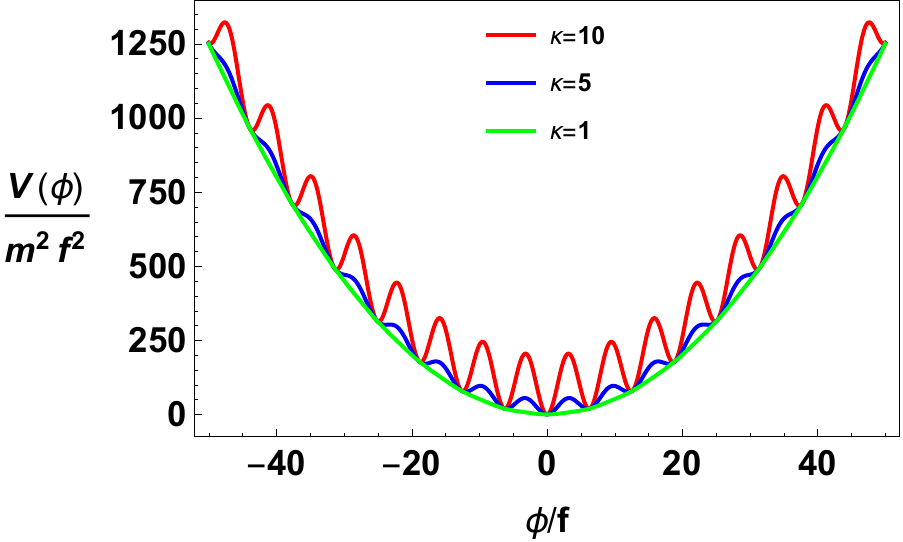}
	\caption{The potential $V(\phi)$ in units of $(mf)^2$, as a function of the field configuration in units of the energy scale $f$. The coloring labels the different values of the parameter $\kappa = \Lambda^2/mf$ considered. \label{V}}
\end{figure}

Motivated by the instanton and monodromy corrections to axion-like potentials we consider~\cite{Kaloper:2008fb,Silverstein:2008sg,Jaeckel:2016qjp}
\begin{eqnarray} \label{eq:Vosc}
    V_{\rm osc}(\phi) = \Lambda^4 \left[1- \cos \left(\frac{\phi}{f}\right)\right] \, ,
\end{eqnarray}
where the quantity $\Lambda$ is a free parameter which controls the height of the perturbations over the quadratic potential, and $f$ controls the field excursion. We show the shape of the potential $V(\phi)$ in Fig.~\ref{V} for different values of the parameter $\kappa=\Lambda^2/(m f)$ which is approximately equal to half the number of metastable minima in the potential.

The field gets trapped in one of the metastable minima whenever the mass correction induced by the potential $V_{\rm osc}(\phi)$ is large~\cite{Jaeckel:2016qjp}, i.e.
\begin{eqnarray}
	M^2 = (1+ \kappa^2)m^2 \gg m^2\, .
\end{eqnarray}

\begin{figure}
	\centering
	\includegraphics[width=8cm]{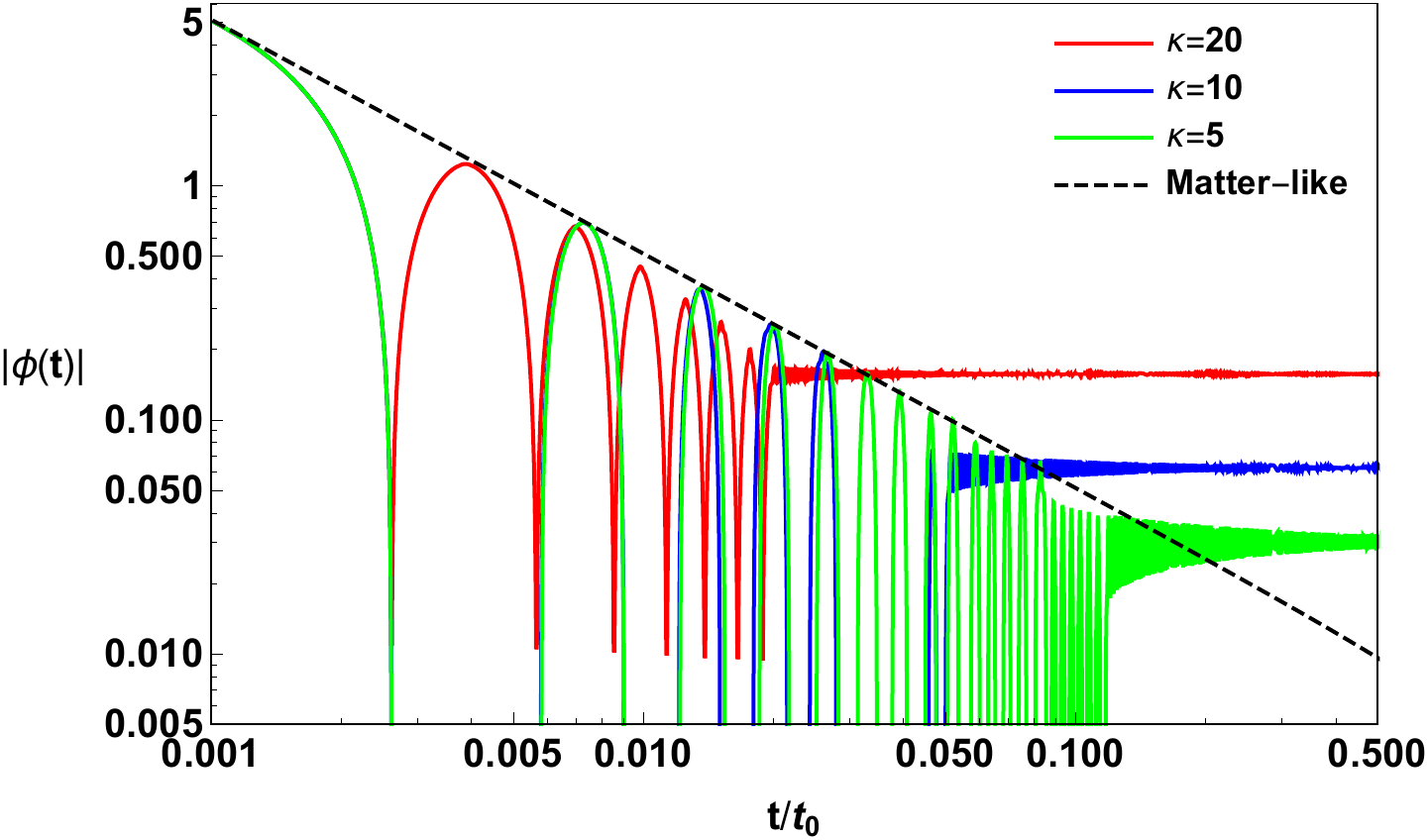}
	\caption{Time evolution of the absolute value of $\phi(t)/f$ as a function of $t/t_*$, for different values of $\kappa$. The dashed lines show the values of $\Lambda^2/m$ for each value of $\kappa$ considered. The dotted line is a fit to a matter-like behavior, for which $\phi \propto a^{-3/2}$. \label{plpht}}
\end{figure}

In Fig.~\ref{plpht} we show the numerical evaluation of Eq.~\ref{eq:KGequation} for the homogeneous mode of $\phi$ using the potential given in Eq.~\ref{eq:Vosc}. We fixed initial conditions such that $\phi(t_i)= \phi_0$ and $\dot{\phi}(t_i)=-m\phi_0$ and used $\kappa \in \{5,10,20\}$.
The evolution of the energy density of the quintessence field shows a transition at a critical time $t_*$. For $t \lesssim t_*$, the scalar field oscillates in the quadratic part of the potential, behaving as a massive scalar field with $\rho_\phi \propto a^{-3}$.  At this stage, the effective equation of state for the quintessential component is $w_\phi \approx 0$ when averaged over many oscillations. For $t \gtrsim t_*$, the oscillatory corrections to the potential become important and the field becomes trapped in one of the metastable minima behaving from then on as a cosmological constant with $w_\phi \simeq-1$. %In Appendix~\ref{sec:initialcondition} we study the dependence of the final value of $\phi$ on the initial conditions.

Let us now investigate the range of parameters where this transition from dark matter to dark energy is efficient. The typical energy of the metastable vacua is $\rho \sim \Lambda^4$. Therefore, in order for the field to be a significant component of the present dark energy density $\Lambda^4 \sim (0.1\, \text{meV})^4$. The condition $\kappa \gg 1$ translates into $\sqrt{m f} \ll 0.1$ meV. For example, for a scalar field which begins oscillating at matter-radiation equality, $m \sim H_\text{eq} \sim 10^{-28}\,$eV, the previous condition requires $f \ll 10^{11}$ GeV. More massive fields would require even smaller energy scales $f$. 

%Apart from ensuring that self-interactions are negligible\footnote{As it was shown in \cite{Jaeckel:2016qjp}, if self-interactions become important these would require a non-perturbative treatment of the perturbations. However, while oscillating around $\phi=0$, the quartic interaction is $\lambda= \Lambda^4/(4f^4)$ which is very small in this context.}
A second important constraint on this class of models arises from considering the tunneling of the $\phi$ field through the potential barrier. In fact, if the field is able to tunnel or jump the barrier, for example due to the inherent quantum fluctuations, it would generate a bubble of a lower energy vacuum. If the tunneling rate is large, $\Gamma_\text{tunneling} \gg H$, bubbles would collide and end precociously the dark energy stage which we have assumed. In~\ref{sec:tunneling} we estimate the tunneling rates and find that if the field inherits an adiabatic spectrum of perturbations from inflation then the Hubble rate at the transition, $H_*$, needs to satisfy
\begin{eqnarray}
    \frac{H_*}{H_\text{eq}}  \gtrsim  7\times 10^{-4} \left(\frac{M}{H_\text{eq}}\right)^{3/5} \, ,
    \label{eq:constrainmodel}
\end{eqnarray}
in order for the field not to tunnel before the present time.

In the data analysis discussed in Sec.~\ref{sec:Method} below, we have not implemented the constraint in Eq.~\eqref{eq:constrainmodel} since it is associated with this particular realization of the model.

\section{Background and perturbations \label{sec:backgroundperturbations}}

In this section we provide the basic equations describing the evolution of both the background and the perturbations of the components that determine the expansion rate of the universe. 

\subsection{Background \label{sec:background}}
We work in the homogeneous, isotropic, and flat Friedmann-Lema\^{i}tre-Robertson-Walker (FLRW) metric with line element
\be
	ds^2 = -dt^2 + a^2(t)d{\bf x}^2,
	\label{eq:metric}
\ee
where $t$ measures the cosmic time, ${\bf x}$ is the vector of spatial coordinates, and $a(t)$ is the scale factor. The Hubble rate is defined as $H = \dot a/a$, where a dot indicates a derivation with respect to cosmic time. In this metric, the Friedmann equations for a homogeneous and isotropic universe read
\bea
	H^2 &\equiv& \(\frac{\dot a }{a}\)^2 = \frac{\rho}{3\MP^2} = \frac{1}{3\MP^2} \sum_i \rho_i, \label{eq:hubble}\\
	\dot H &=& -\frac{1}{2\MP^2} \sum_i \(1+w_i\)\,\rho_i, \label{eq:hubble2}
\eea
where $\MP = \(8\pi G_N\)^{-1/2}$ is the reduced Planck mass given in terms of Newton's constant $G_N$. In the fluid description, we consider the energy density $\rho_i$ of the $i$-th component, with the index $i$ running over the set
\be
	i \in \mathcal{S}= \{b, c, R, \Lambda, \phi\},
	\label{eq:set1}
\ee
corresponding to baryons $(b)$, dark matter $(c)$, radiation $(R)$, cosmological constant $(\Lambda)$, and the quintessence field $(\phi)$. Each component of the energy density satisfies the non-interacting continuity equation
\be
	\dot\rho_i + 3H(p_i+\rho_i) = 0,
	\label{eq:continuity_i}
\ee
with pressure $p_i$ and equation of state $w_i = p_i / \rho_i$. We have assumed that from the period at which recombination occurs until present time, the different components have contributed to the expansion rate of the universe with $w_b = w_c = 0$ for matter, $w_R = 1/3$ for radiation, $w_\Lambda = -1$ for the cosmological constant. In general, the equation of state for the quintessence field reads
\be
	w_\phi = \frac{p_\phi}{\rho_\phi} = \frac{\frac{1}{2}\dot\phi^2 - V(\phi)}{\frac{1}{2}\dot\phi^2 + V(\phi)}\,,
	\label{eq:wphi}
\ee
so that it evolves with time within the range $w_\phi \in \[-1,1\]$. In the numerical analysis, we effectively describe the equation of state for the quintessence fluid through Eq.~\eqref{eq:weff}, setting $w_\phi = w_{\rm \phi eff}(a)$. The pressure and the energy density in the quintessence field are respectively the numerator and the denominator in Eq.~\eqref{eq:wphi}. The sum of all continuity equations can be rephrased to express the conservation law for the total energy density $\rho \equiv \sum_i \rho_i$, as
\bea
	\dot\rho &=& -3H(1 + w_{\rm eff})\rho\,, \label{eq:continuity_all}\\
	w_{\rm eff} &\equiv& \frac{\sum_i w_i \rho_i}{\sum_i \rho_i}\,.
\eea
For future convenience, we define the fractional energy density contribution for each fluid at present time $\Omega_i \equiv \rho_i/\rho_{\rm crit}$, in units of the present critical energy density $\rho_{\rm crit} = 3\MP^2H_0^2$. We also define the rescaled Hubble rate $h = H_0/100{\rm \,km \,s^{-1}\,Mpc^{-1}}$.

\begin{figure}[h!]
\begin{center}
	\includegraphics[width=0.7\linewidth]{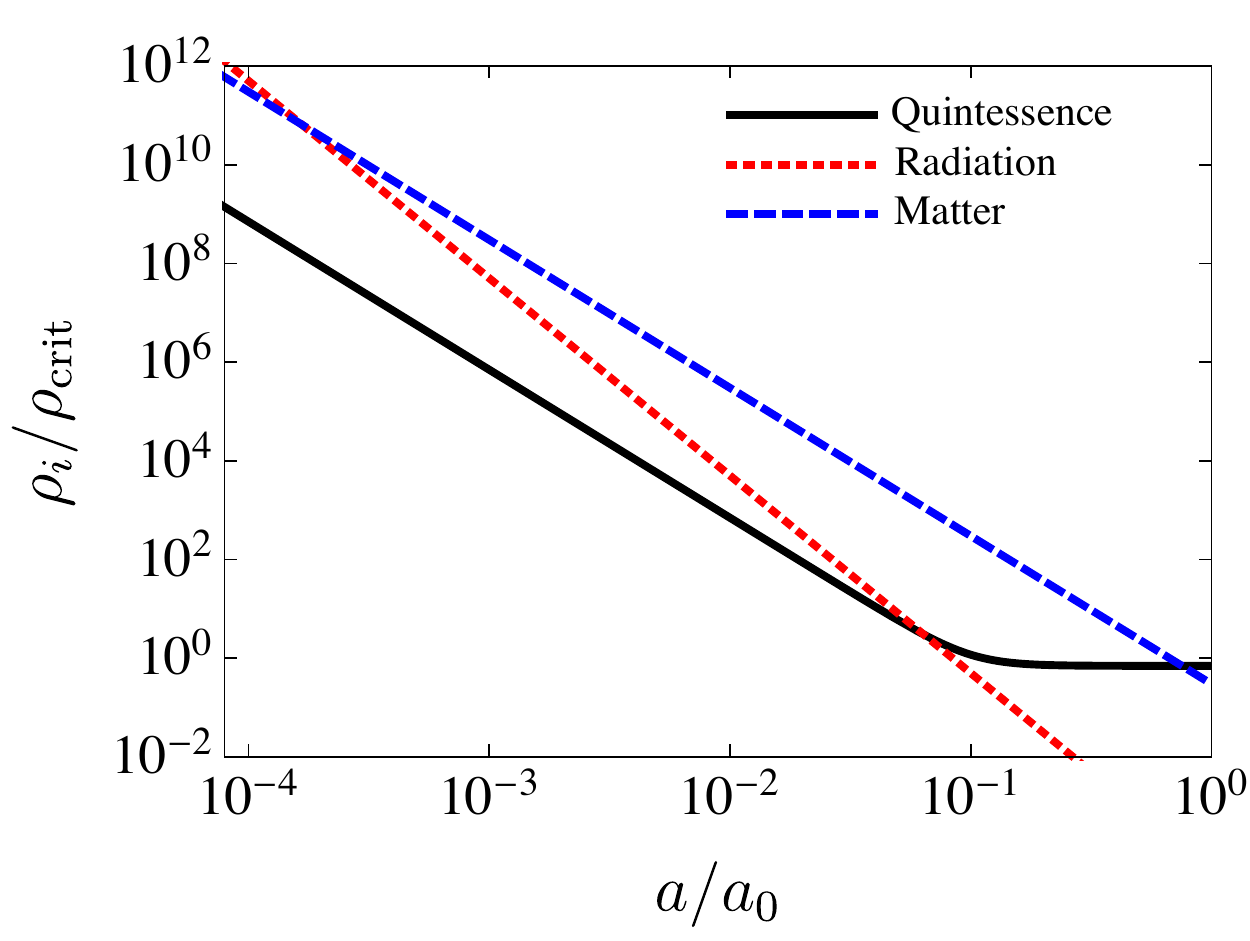}
	\caption{Evolution of the energy density in the quintessence field (black solid line), radiation (red dotted line), and  matter (blue dashed line), as a function of the scale factor $a$. The quintessence field is described by the equation of state in Eq.~\eqref{eq:weff} with $\Delta = 0.5$, $a_* = 10^{-1}$, $f_L= 0$, and $w_{\phi 0} = -1$, see the text for additional detail. For each species $i$, the corresponding energy density $\rho_i$ is measured in units of the present critical density $\rho_{\rm crit} = 3\MP^2H_0^2$.}
	\label{fig:hubblesolve}
\end{center}
\end{figure}

\begin{figure}[h!]
\begin{center}
	\includegraphics[width=0.7\linewidth]{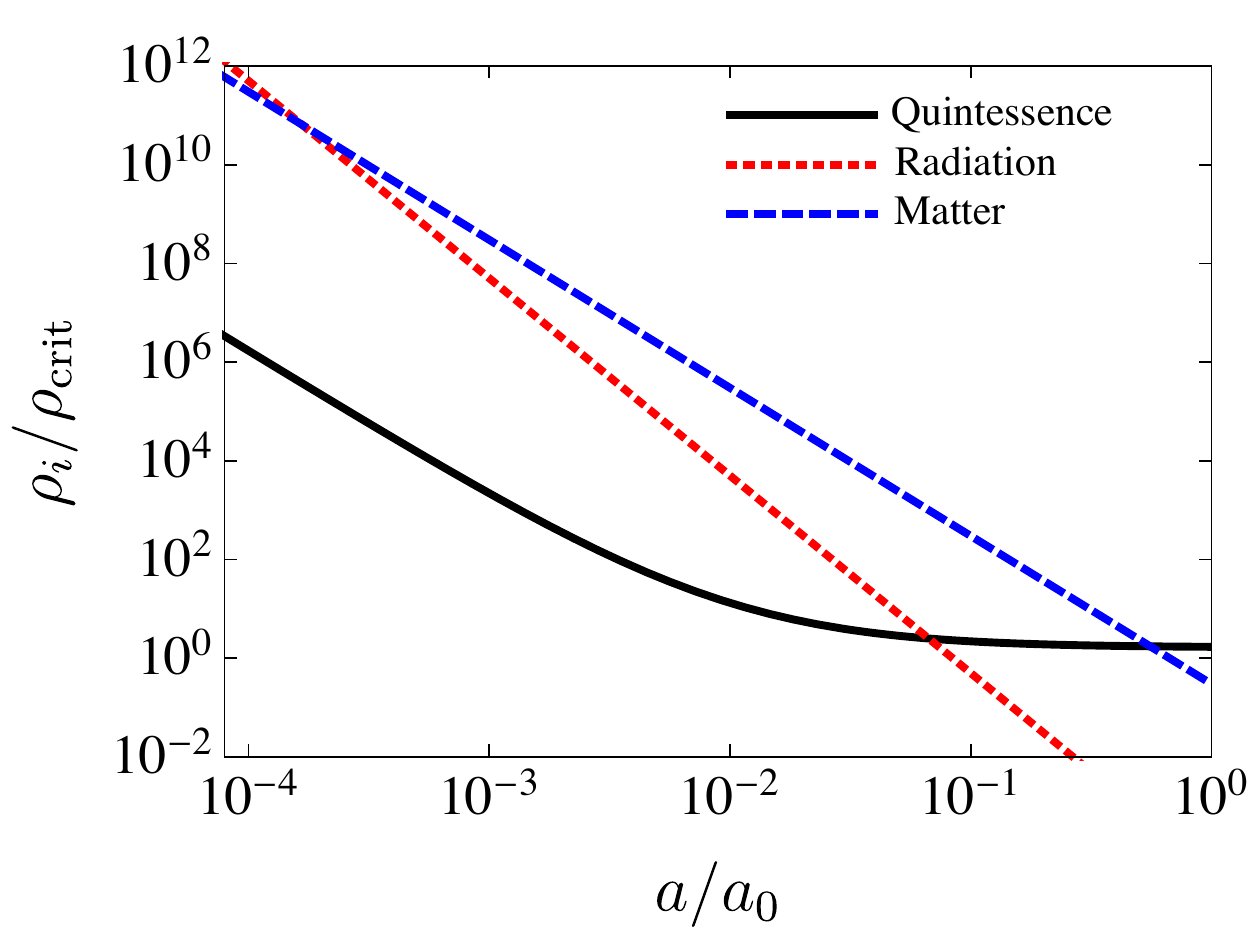}
	\caption{Same as Fig.~\ref{fig:hubblesolve}, with the parameters $\Delta = 2$, $a_* = 10^{-2}$, $f_L= -1$, and $w_{\phi 0} = -1$, see the text for additional detail.}
	\label{fig:hubblesolve2}
\end{center}
\end{figure}

In Fig.~\ref{fig:hubblesolve} we plot the energy density in the quintessence field (black solid line), matter (blue dashed line) and radiation (red dotted line) as a function of the scale factor $a$. In order to plot the evolution of the various energy contents, we have accounted for the constraint that expresses the energy content in matter today as
\be
	\Omega_M \equiv \Omega_b + \Omega_c = 1-\Omega_\phi - \Omega_R - \Omega_\Lambda.
	\label{eq:consistency}
\ee
In order to plot the results in Fig.~\ref{fig:hubblesolve} we have also set to zero the cosmological constant $\Omega_\Lambda = 0$, and we have fixed the present contribution to the total energy budget of the quintessence field and of radiation respectively to $\Omega_\phi = 0.7$ and $\Omega_R = 5\times 10^{-5}$. These choices are not implemented further in our data analyses of Sec.~\ref{sec:Results}, where only the consistency relation in Eq.~\eqref{eq:consistency} takes place. The energy density in the quintessence field in Fig.~\ref{fig:hubblesolve} is described by Eq.~\eqref{eq:energydensity} in which, for illustrative purpose, we have set $\Delta = 0.5$, $a_* = 10^{-1}$, $f_L=0$, and $w_{\phi 0} = -1$, so that the quintessence field tracks the dark matter behaviour from recombination upon transitioning to a behaviour like that of a cosmological constant around the scale factor $a_*$. The black line in Fig.~\ref{fig:hubblesolve} describing the quintessence field mimics the behaviour of the particle physics model shown in Fig.~\ref{plpht} as the solution to the Klein-Gordon Eq.~\eqref{eq:KGequation}.
A different choice of the parameters describing the quintessence field leads to a modified cosmological history. Fig.~\ref{fig:hubblesolve2} shows the evolution of the energy densities for a non-zero and negative cosmological constant $\Omega_\Lambda = -1$, so that we have fixed the present contribution in the dark energy sector (quintessence plus cosmological constant) as $\Omega_\phi + \Omega_\Lambda = 0.7$. Additionally, we have set $\Delta = 2$, $a_* = 10^{-2}$, $f_L=-1$, and $w_{\phi 0} = -1$. These numerical values are chosen for illustrative purpose and do not enter the MCMC analysis described in Sec.~\ref{sec:Results}. The effect of a larger width $\Delta = 2$ is expressed in a prolonged transition period of the quintessence field in Fig.~\ref{fig:hubblesolve2} (solid black line) with respect to what obtained in Fig.~\ref{fig:hubblesolve} with $\Delta = 2$. In addition, the transition in the second example occurs earlier since $a_*$ is smaller, and the quintessence field today reaches a larger value since it has to balance out the negative contribution of the cosmological constant in order to reproduced the observed accelerated expansion rate today.

\subsection{Perturbations \label{sec:perturbations}}

The linear perturbations in the quintessence field evolve according to the perturbed Klein-Gordon Eq.~\eqref{eq:KGequation}. We consider the fluid counterpart of the Klein-Gordon equation by averaging over the field oscillations, since the oscillations of the scalar field occur on a much shorter timescale than a Hubble time~\cite{Turner:1983he}. Here, the expressions describing perturbations are expressed in the synchronous gauge~\cite{Mukhanov:1990me, Ma:1995ey, Hu:1998kj}, in which the perturbations over the FLRW metric in Eq.~\eqref{eq:metric} depend on the tensor $h_{ij}$ in the conformal time $\tau$ as
\be
	ds^2 = a^2(\tau)\[-d\tau^2 + \(\delta_{ij} + h_{ij}\)dx^idx^j\].
\ee
In the synchronous gauge, the equations governing the evolution of density and bulk velocity perturbations in the quintessence field can be written in terms of fluid variables as~\cite{Weller:2003hw}
\begin{eqnarray}
    \delta_\phi' &=& -3\HH\(c_{s, \phi}^2 - w_\phi\)\delta_\phi -9\HH^2 \(c_{s, \phi}^2 -c_{a, \phi}^2\)\frac{\theta_\phi}{k^2} - \nonumber\\
    && -\theta_\phi - 3(1+w_\phi)\,h',\\
    \theta_\phi' &=& -\(1-3\(c_{s, \phi}^2 - c_{a, \phi}^2 + w_\phi\)\)\HH \theta_\phi + c_{s,\phi}^2k^2\delta_\phi,
\end{eqnarray}
where a prime indicates a derivative with respect to conformal time $\tau$, with $\HH \equiv a'/a$. For the quintessence model we implemented the effective sound speed $c_{s, \phi}^2 \equiv \partial p_\phi/\partial \rho_\phi$ which might differ from one. We have also defined the adiabatic sound speed, which depends only on background quantities, as
\be
	c_{a, \phi}^2 \equiv \frac{p'_\phi}{\rho'_\phi} = w_\phi - \frac{w_\phi'}{3\HH\(1+w_\phi\)},
\ee
while we defined $\theta_\phi = (1+w_\phi)kv_\phi$, in terms of the bulk velocity $v_\phi$.

We have included the effects arising from the interacting fluids~\cite{Malik:2004tf, Malik:2008im, Gavela:2009cy, Gavela:2010tm, Visinelli:2011jy, Visinelli:2014qla, Visinelli:2016rhn, Freese:2017ace, Yang:2018pej} in the synchronous gauge. The momentum transfer is zero in the rest frame of the dark matter component. We adopt adiabatic initial conditions for the quintessence component~\cite{Valiviita:2008iv, He:2008si, Gavela:2009cy, Gavela:2010tm}, as well as for all the other constituents that show up in the set in Eq.~\eqref{eq:set1}~\cite{Ma:1995ey}. The evolution of baryons, cold dark matter, radiation, and neutrinos are accounted for by the Boltzmann scheme we introduce in Sec.~\ref{sec:Method}.

% For this, we have set the equations of state for both baryons and CDM $w_b = w_c = 0$, while $w_\phi$ has been parametrised as in Eq.~\eqref{eq:weff}. We also set the effective sound speed for each component $c_{s, b}^2 = c_{s, c}^2 = 0$~\cite{Majerotto:2009np}, \RZF{Isn't $c_{s,b}^2, w_b \neq 0$?} while for the quintessence field we set $c_{s,\phi}^2 = {\rm abs}(w_\phi)$. 
%The expressions for the perturbations have been specialized to our model in Appendix~\ref{appendix2}.

\section{Method}
\label{sec:Method}

To better address the change in the cosmological analysis due to the presence of the additional quintessence field, we have modified the publicly available numerical Boltzmann solver code \texttt{CAMB}~\cite{Lewis:1999bs} by implementing the relevant equations for our purpose. We perform a Monte Carlo Markov Chain (MCMC) analysis using the August 2018 version of the publicly available package \texttt{CosmoMC}~\cite{Lewis:2002ah}, modified to include the additional parameters that define the theory, namely the scale factor $a_*$, the fraction of the energy density in quintessence field $f_L = \rho_{\phi,0} / \rho_\Lambda$ and the duration of the transition $\Delta$. \texttt{CosmoMC} includes the support for the \textit{Planck} data release 2015 Likelihood Code~\cite{Aghanim:2015xee} (see \url{http://cosmologist.info/cosmomc/}) and implements an efficient sampling by using the fast/slow parameter decorrelations~\cite{Lewis:2013hha}. When performing the MCMC analysis, we have taken into account the BBN consistency to calculate the primordial abundances of helium and deuterium based on $\Omega_b h^2$ and $\Delta N_{\rm eff}$. However, in the code we have not included any extra contribution to $\Delta N_{\rm eff}$, while $\Omega_b h^2$ is modified with respect to the $\Lambda$CDM value by the presence of the quintessence field as an extra DM contribution during recombination.

As a baseline, we consider a total of seven parameters varying independently: the six parameters of the standard $\Lambda$CDM model (the baryon $\Omega_{\rm b}h^2$ and cold dark matter $\Omega_{\rm c}h^2$ energy densities, the ratio between the sound horizon and the angular diameter distance $\theta_{\rm{MC}}$, the reionization optical depth $\tau$, the amplitude $\log_{10}(10^{10}A_{\rm s})$ and the scalar spectral index $n_{\rm s}$ of the primordial scalar spectrum, and $\log_{10}(a_*)$). As a second step, in order to test the robustness of our assumptions, we add two more parameters defining the model we are exploring here: $f_L$ and $\Delta$. Finally, we consider some standard extension of the $\Lambda$CDM model like the dark energy equation of state $w_{\phi 0}$ or the neutrino sector parameters, i.e. the total neutrino mass $\Sigma m_{\nu}$ and the neutrino effective number $N_{\rm eff}$. In all our cases we also vary the foreground parameters as described in Refs.~\cite{Aghanim:2015xee, Ade:2015xua}. All the parameters considered in this paper are varying in a range of flat conservative priors listed in Table~\ref{priors}.

\begin{table}
\begin{center}
\renewcommand{\arraystretch}{1.4}
\begin{tabular}{|c@{\hspace{1 cm}}|@{\hspace{1 cm}} c|}
\hline
\textbf{Parameter}                    & \textbf{Prior}\\
\hline\hline
$\Omega_{\rm b} h^2$         & $[0.005\,,\,0.1]$\\
$\Omega_{\rm c} h^2$       & $[0.001\,,\,0.99]$\\
$\theta_{\rm {MC}}$             & $[0.5\,,\,10]$\\
$\tau$                       & $[0.01\,,\,0.8]$\\
$\log_{10}(10^{10}A_{s})$         & $[2\,,\,4]$\\
$n_s$                        & $[0.8\,,\, 1.2]$\\
$f_L$ & $[0\,,\,1]$\\
$\log(a_*)$ & $[-4\,,\,0]$\\
$\Delta$ & $[0.1 \,,\,1]$\\ 
$w_{\phi 0}$ & $[-1 \,,\,1]$\\ 
$\Sigma m_{\nu}$ & $[0 \,,\,5]$\\ 
$N_{\rm eff}$ & $[0.05 \,,\,10]$\\ 
\hline
\end{tabular}
\end{center}
\caption{Flat priors on the cosmological parameters assumed in this paper.}
\label{priors}
\end{table}

The publicly available datasets we analysed in this work are:
\begin{itemize}

\item \textit{Planck}: the full range of the Cosmic Microwave Background measurements from \textit{Planck} 2015, which include the temperature and polarization power spectra data \cite{Aghanim:2015xee}. 

\item Lensing: the 2018 \textit{Planck} measurements of the CMB lensing potential power spectrum $C^{\phi\phi}_\ell$ \cite{Aghanim:2018eyx}.

\item BAO: the baryon acoustic oscillations distance measurements given by the 6dFGS~\cite{Beutler:2011hx}, SDSS-MGS~\cite{Ross:2014qpa}, and 
BOSS DR12~\cite{Alam:2016hwk} surveys, as 
adopted by the \textit{Planck} collaboration~\cite{Aghanim:2018eyx}.

\item DES: the first-year of the Dark Energy Survey lensing cosmic shear measurements~\cite{Troxel:2017xyo, Abbott:2017wau, Krause:2017ekm}, as implemented by the \textit{Planck} collaboration~\cite{Aghanim:2018eyx}.

\item Pantheon:  the most latest compilation of Supernovae Type Ia data comprising 1048 data points~\cite{Scolnic:2017caz}. 

\end{itemize}

We decided in this work not to use the gaussian prior on the Hubble constant as measured by SH0ES~\cite{Riess:2019cxk}, because, as we will see in the next section, the Hubble constant obtained within this model is always in tension with SH0ES at more than $3\sigma$.

\begin{figure}[tb]
	\includegraphics[width=0.5\linewidth]{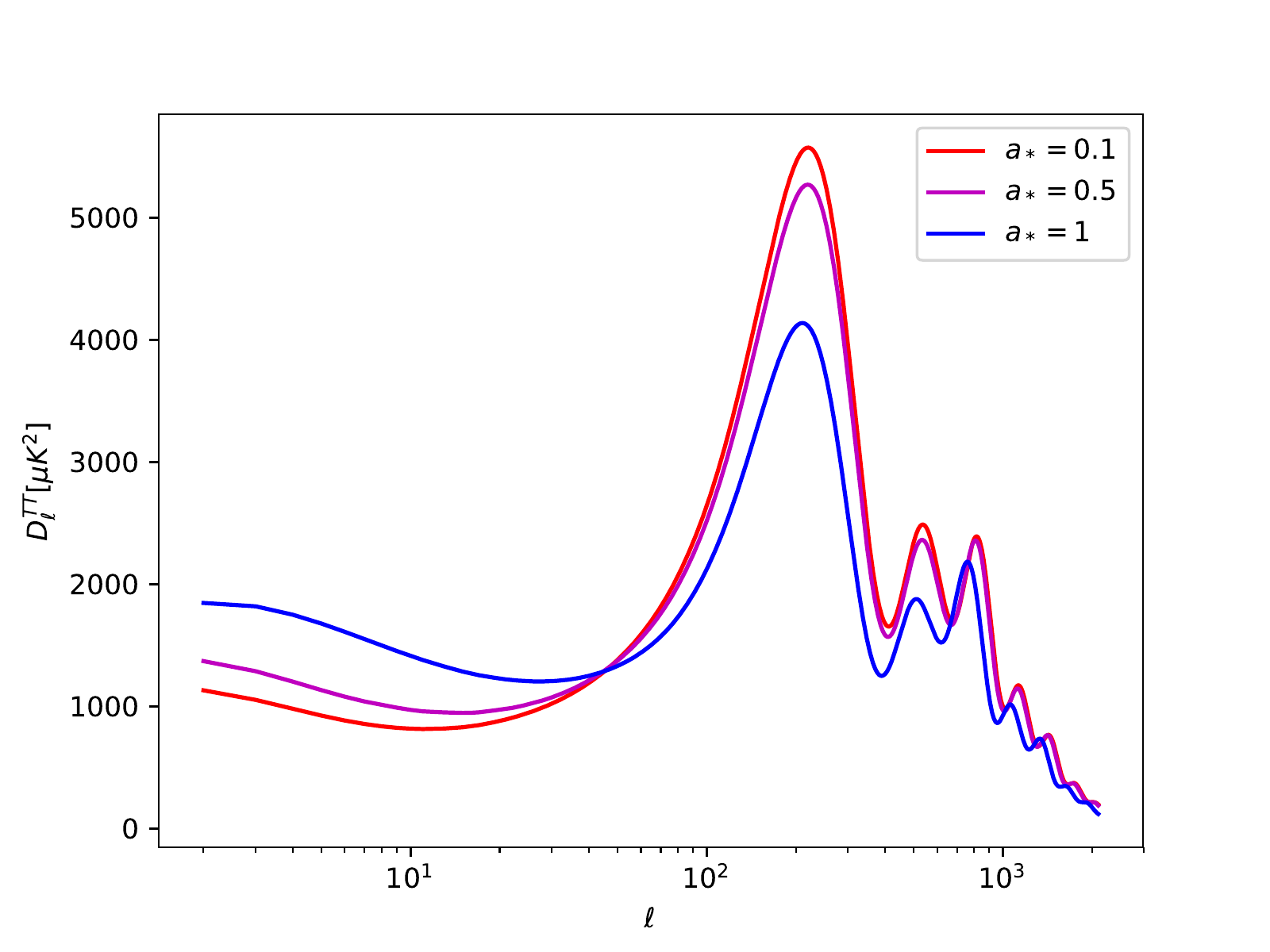}
	\includegraphics[width=0.5\linewidth]{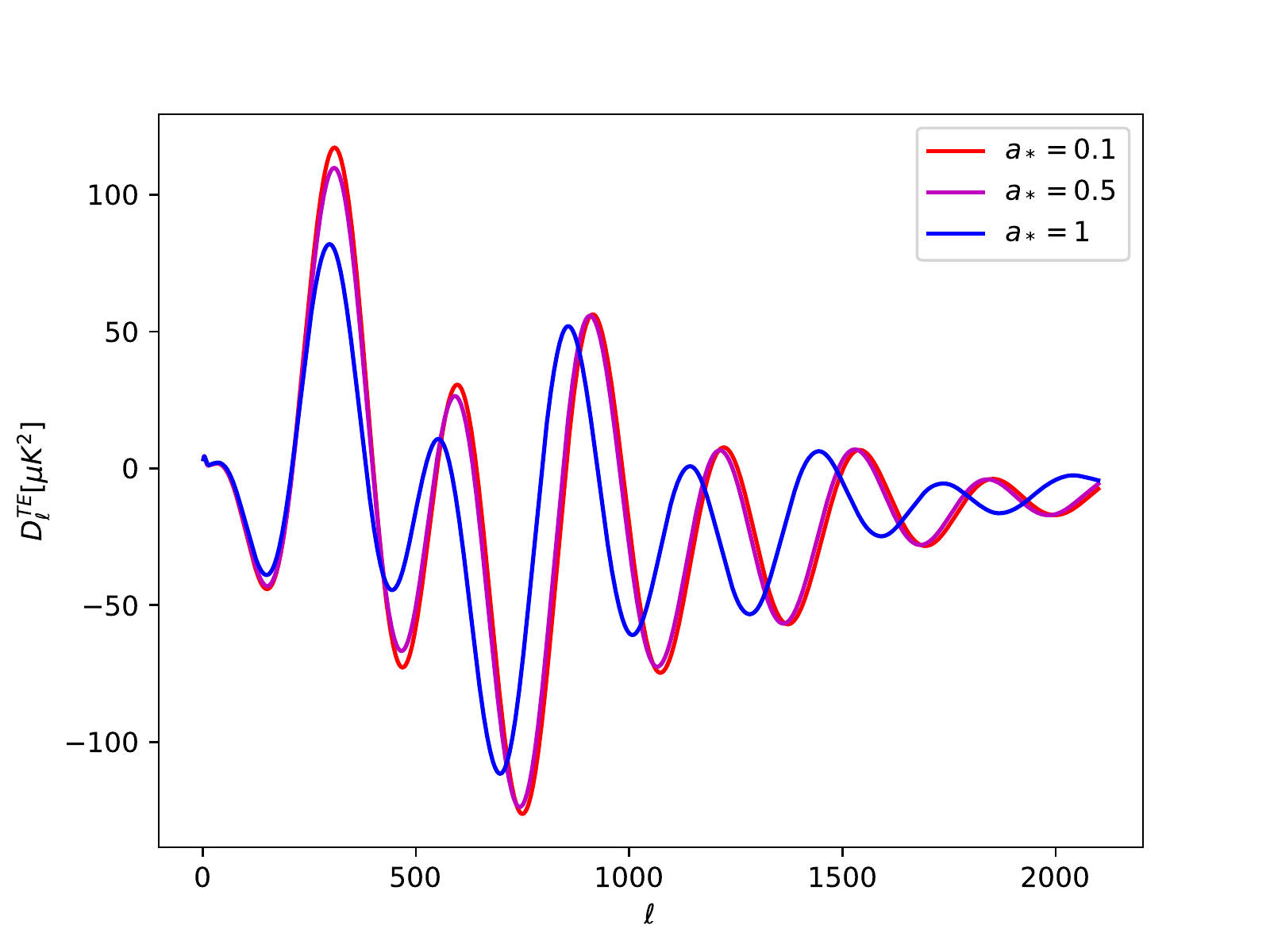}
	\includegraphics[width=0.5\linewidth]{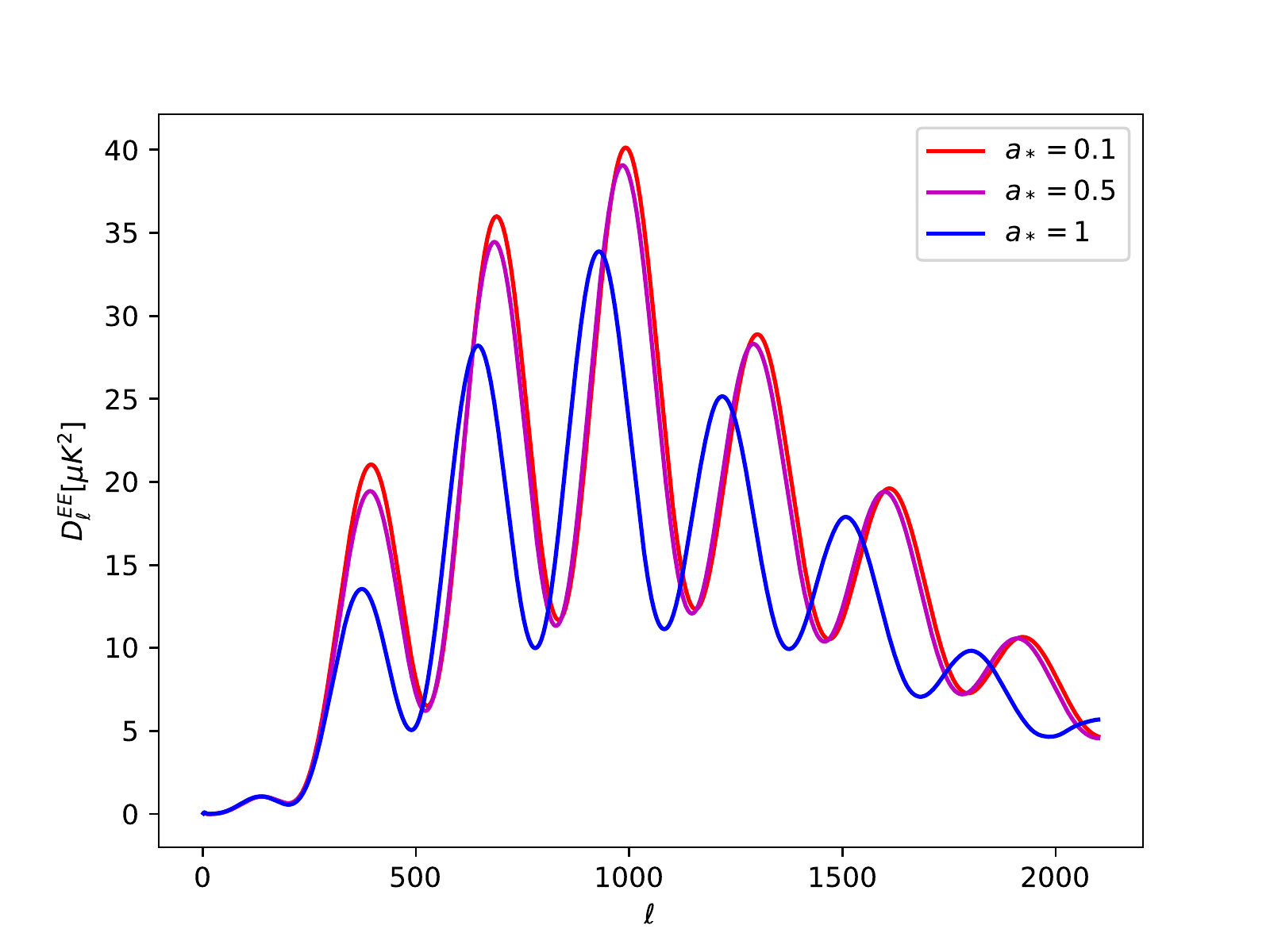}
	\caption{Temperature and polarization power spectra obtained by varying $a_*$ and fixing $f_L=0.7$ and $\Delta=0.2$. By increasing $a_*$ there is a shift of the peaks in the damping tail of the temperature spectrum and in the polarization spectra towards lower multipoles, a suppression of the amplitude of the peaks, and an enhancement of the low-$\ell$ tail in TT.}
	\label{fig:a*}
\end{figure}

\begin{figure}[tb]
	\includegraphics[width=0.5\linewidth]{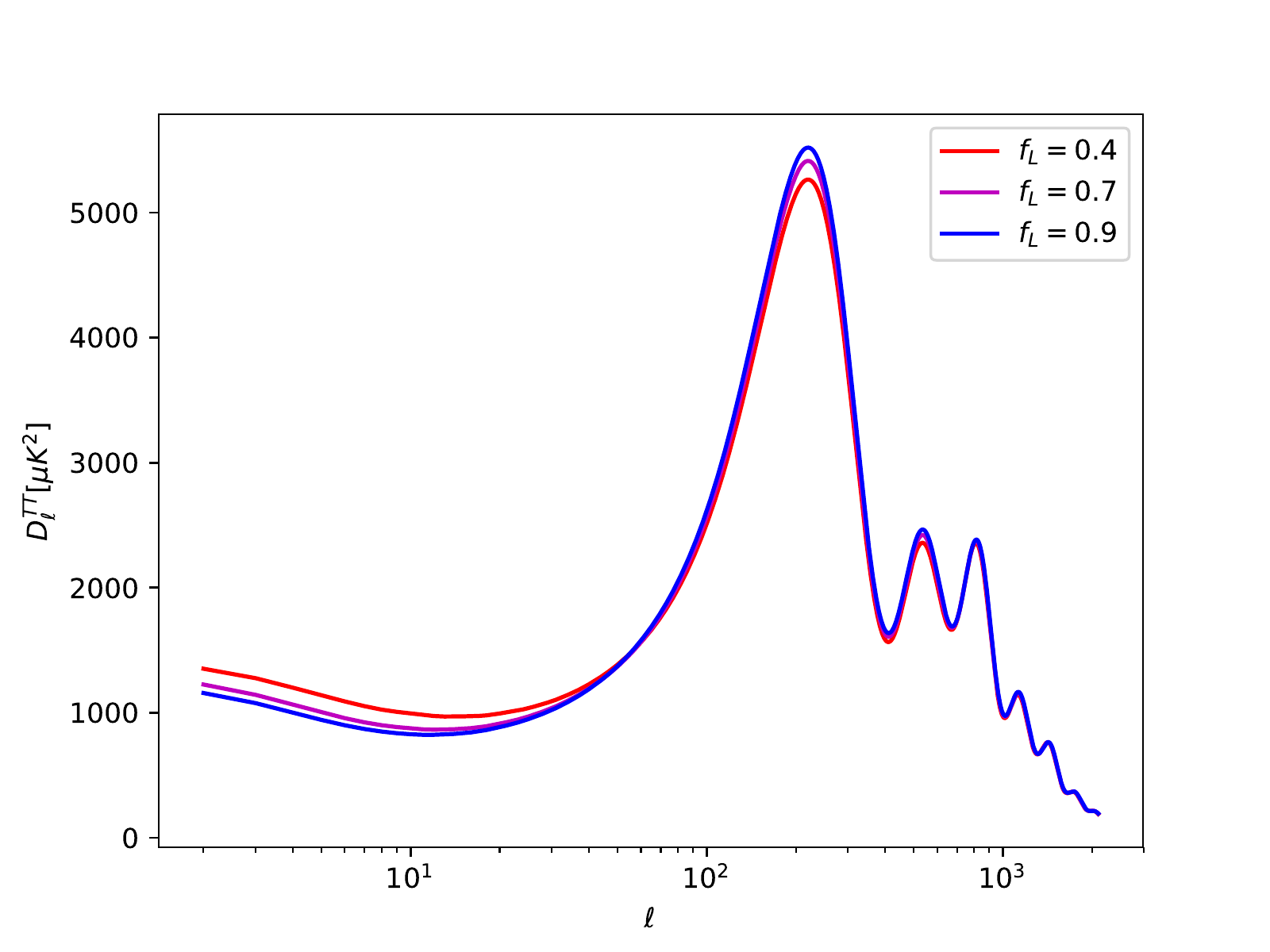}
	\includegraphics[width=0.5\linewidth]{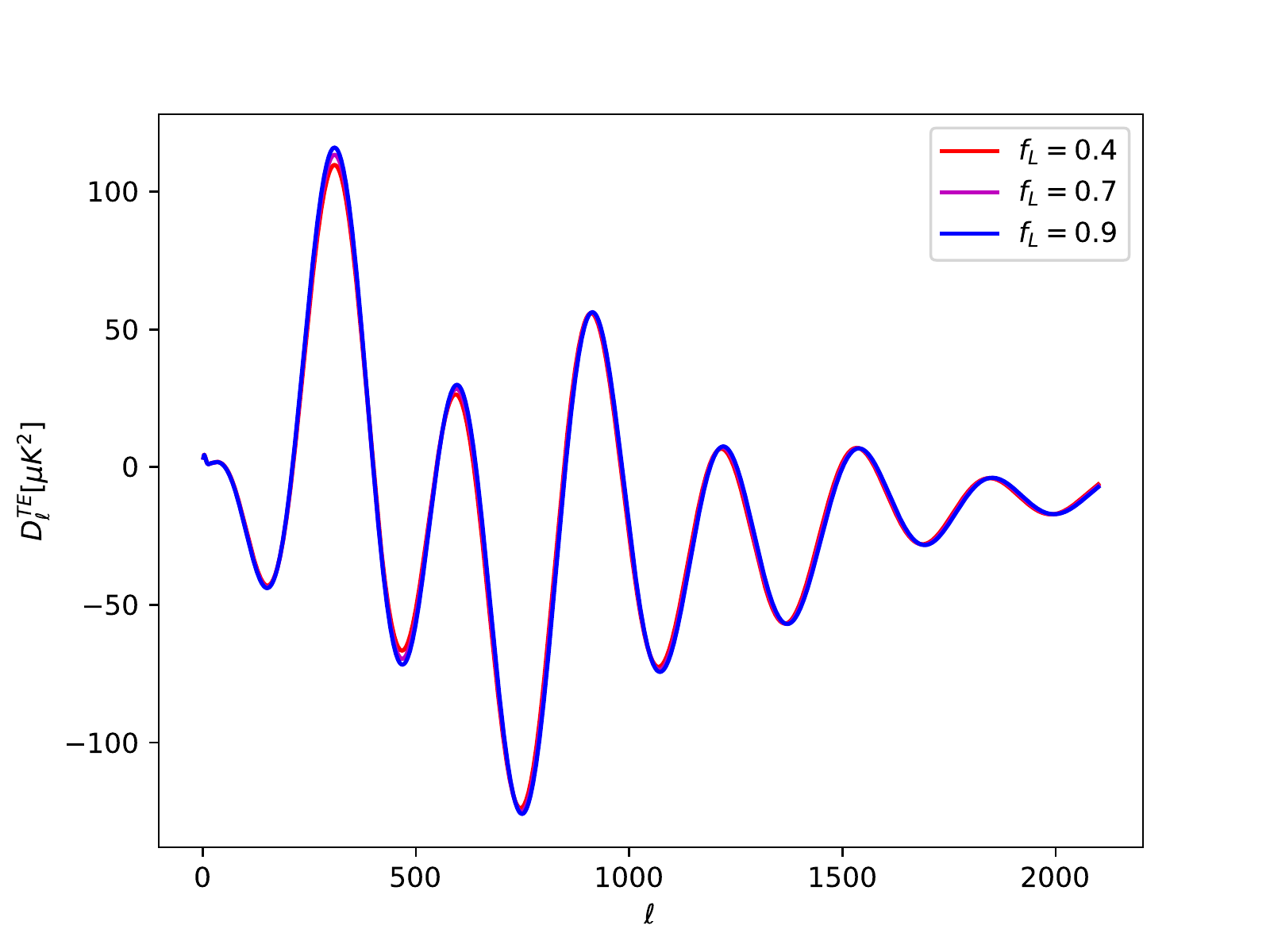}
	\includegraphics[width=0.5\linewidth]{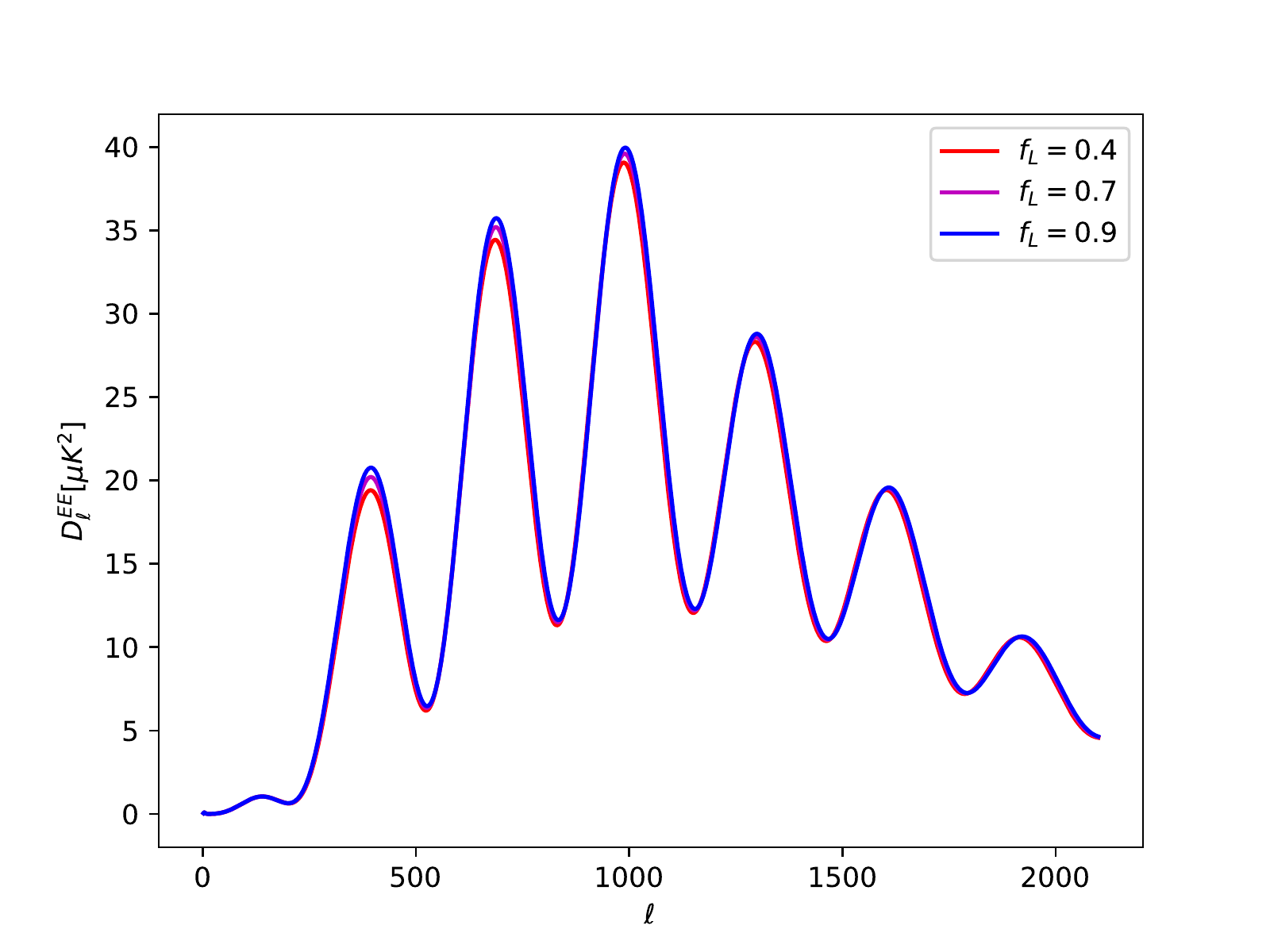}
	\caption{Temperature and polarization power spectra obtained by varying $f_L$ and fixing $a_*=0.4$ and $\Delta=0.2$. The effect of $f_L$ is very small.}
	\label{fig:fl}
\end{figure}

\begin{figure}[tb]
	\includegraphics[width=0.5\linewidth]{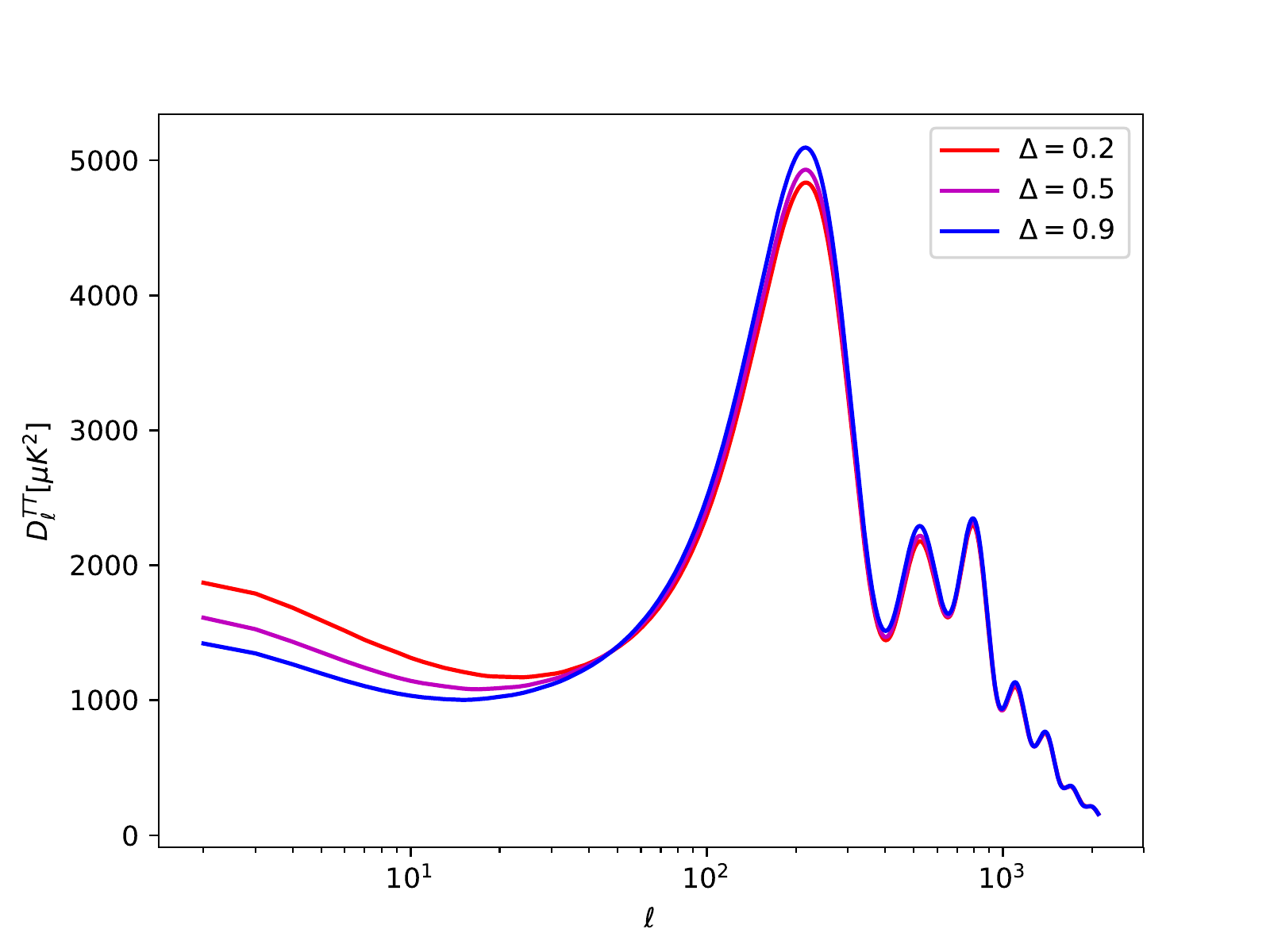}
	\includegraphics[width=0.5\linewidth]{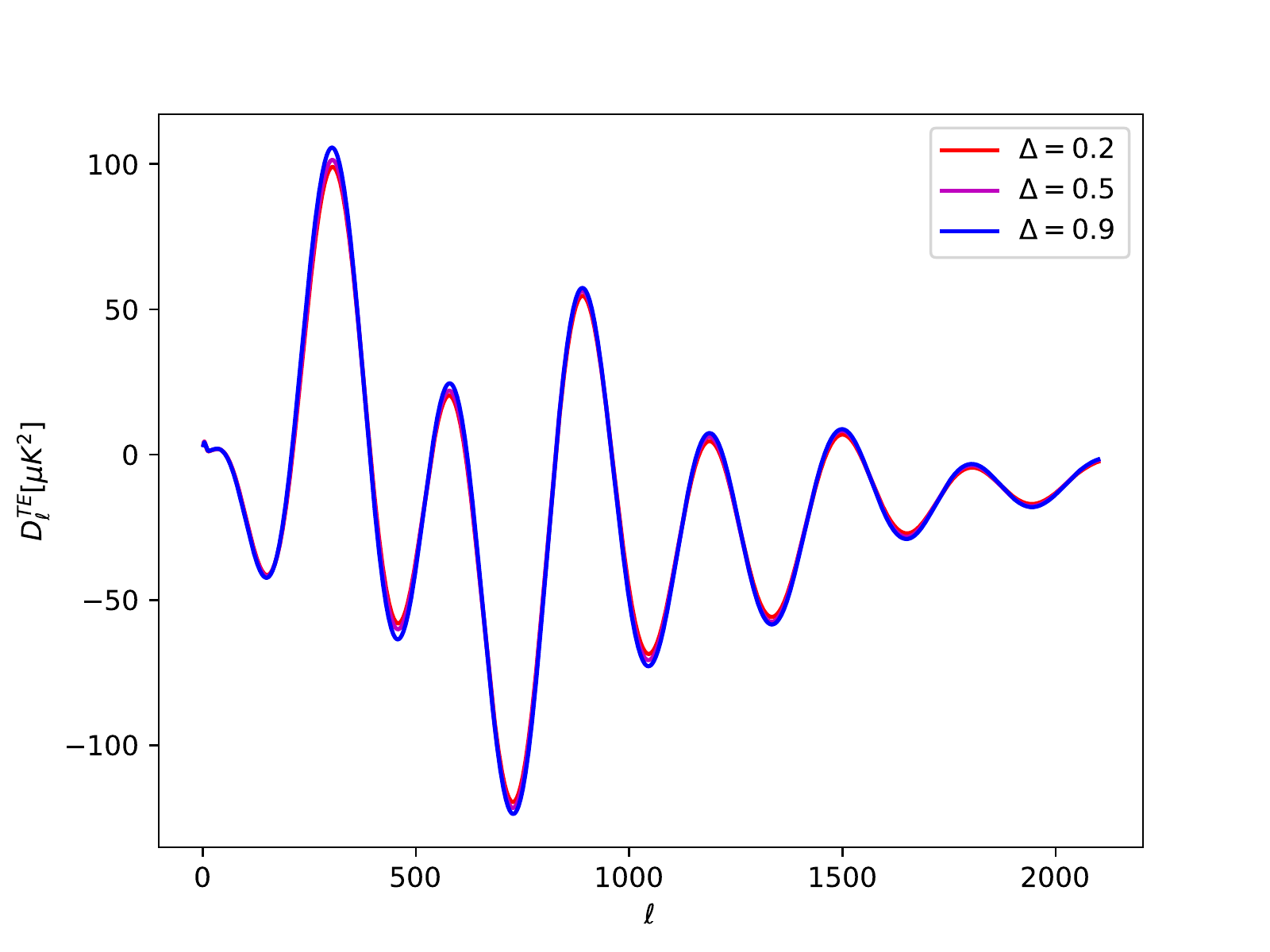}
	\includegraphics[width=0.5\linewidth]{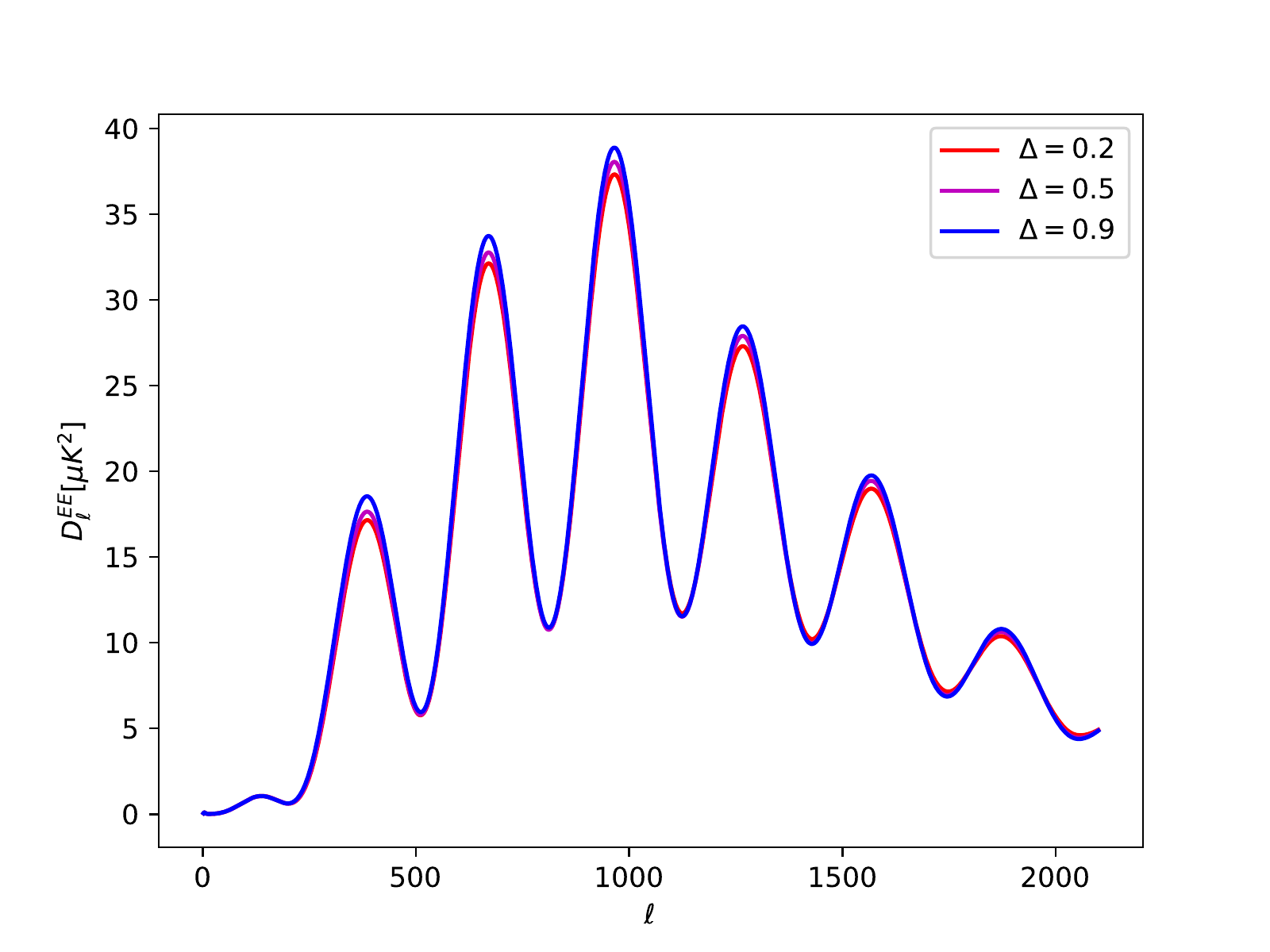}
	\caption{Temperature and polarization power spectra obtained by varying $\Delta$ and fixing $a_*=0.7$ and $f_L=0.7$. By increasing $\Delta$ the main effect is in the suppression of the low-$\ell$ tail in TT.}
	\label{fig:delta}
\end{figure}

We can see the qualitative effect of varying the parameters of the model ($a_*$, $f_L$ and $\Delta$) on the temperature and polarization power spectra in Figs.~\ref{fig:a*},~\ref{fig:fl} and~\ref{fig:delta}. In Fig.~\ref{fig:a*} we observe that by increasing $a_*$ we have a shift of the peaks in the damping tail of the temperature spectrum and in the polarization spectra towards lower multipoles, a suppression of the amplitude of the peaks, and an enhancement of the low-$\ell$ tail in TT. For values of $a_*<10^{-1}$ the spectra are almost indistinguishable, so we don't show them in the plots. In Fig.~\ref{fig:fl} we can see that the effect of $f_L$ is very small, while in Fig.~\ref{fig:delta} we have that increasing $\Delta$ the main effect is in the suppression of the low-$\ell$ tail in TT. The effects of decreasing $f_L$ and $\Delta$ are similar because when either decrease, the amount of dark matter at recombination increases.

\section{Results}
\label{sec:Results}

\begin{figure}[h!]
\includegraphics[width=0.5\textwidth]{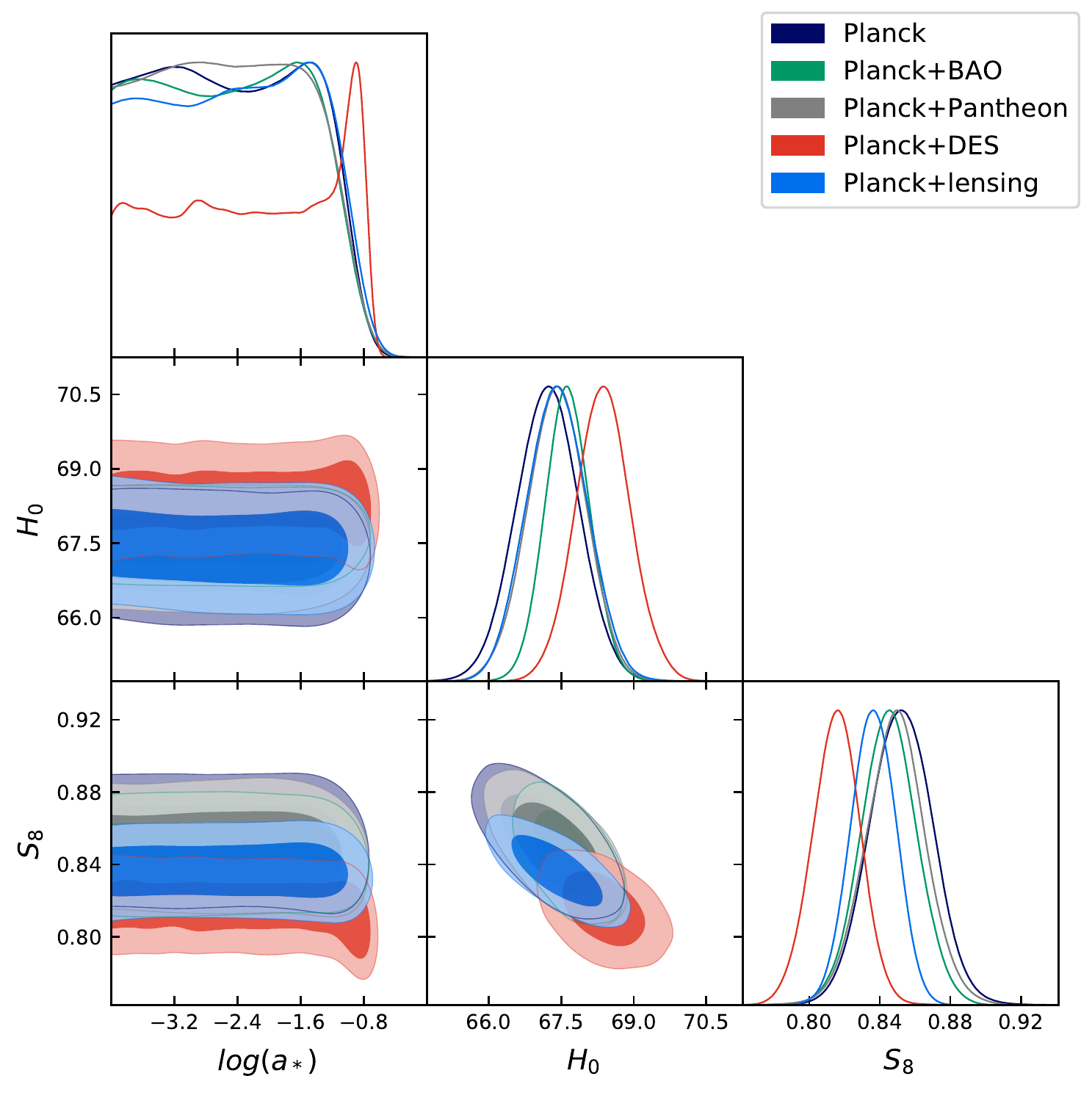}
\caption{2-D contours at 68$\%$~CL and 95$\%$~CL, and 1-D posterior distributions on some selected cosmological parameters, obtained by fixing $f_L=1$ and $\Delta=0.5$.}
\label{Fig:a*}
\end{figure}

\begin{figure}[h!]
\includegraphics[width=0.5\textwidth]{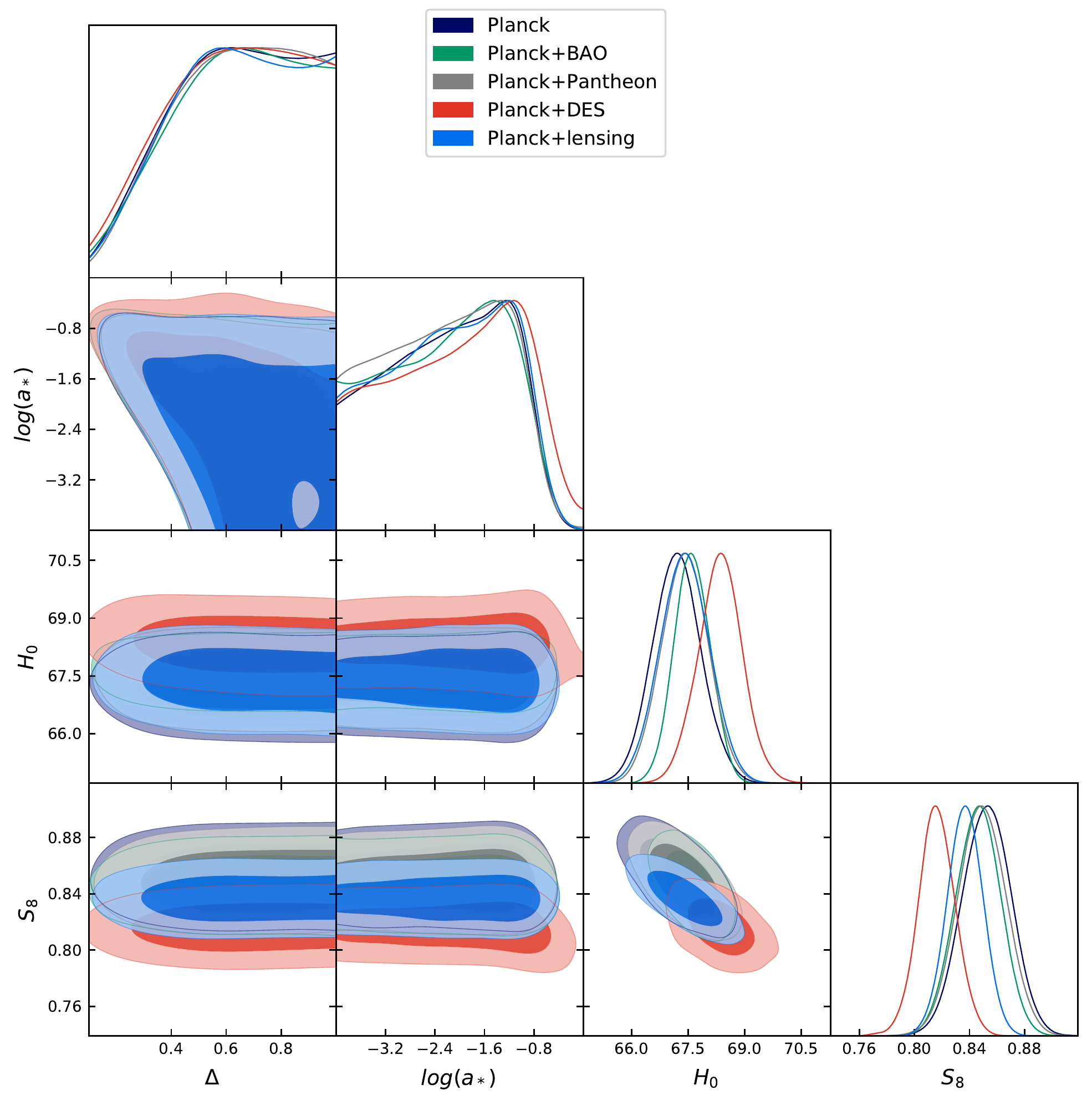}
\caption{2-D contours at 68$\%$~CL and 95$\%$~CL, and 1-D posterior distributions on some selected cosmological parameters, obtained by varying also $f_L$ and $\Delta$.}
\label{Fig:delta}
\end{figure}

\begin{figure}
\includegraphics[width=0.5\textwidth]{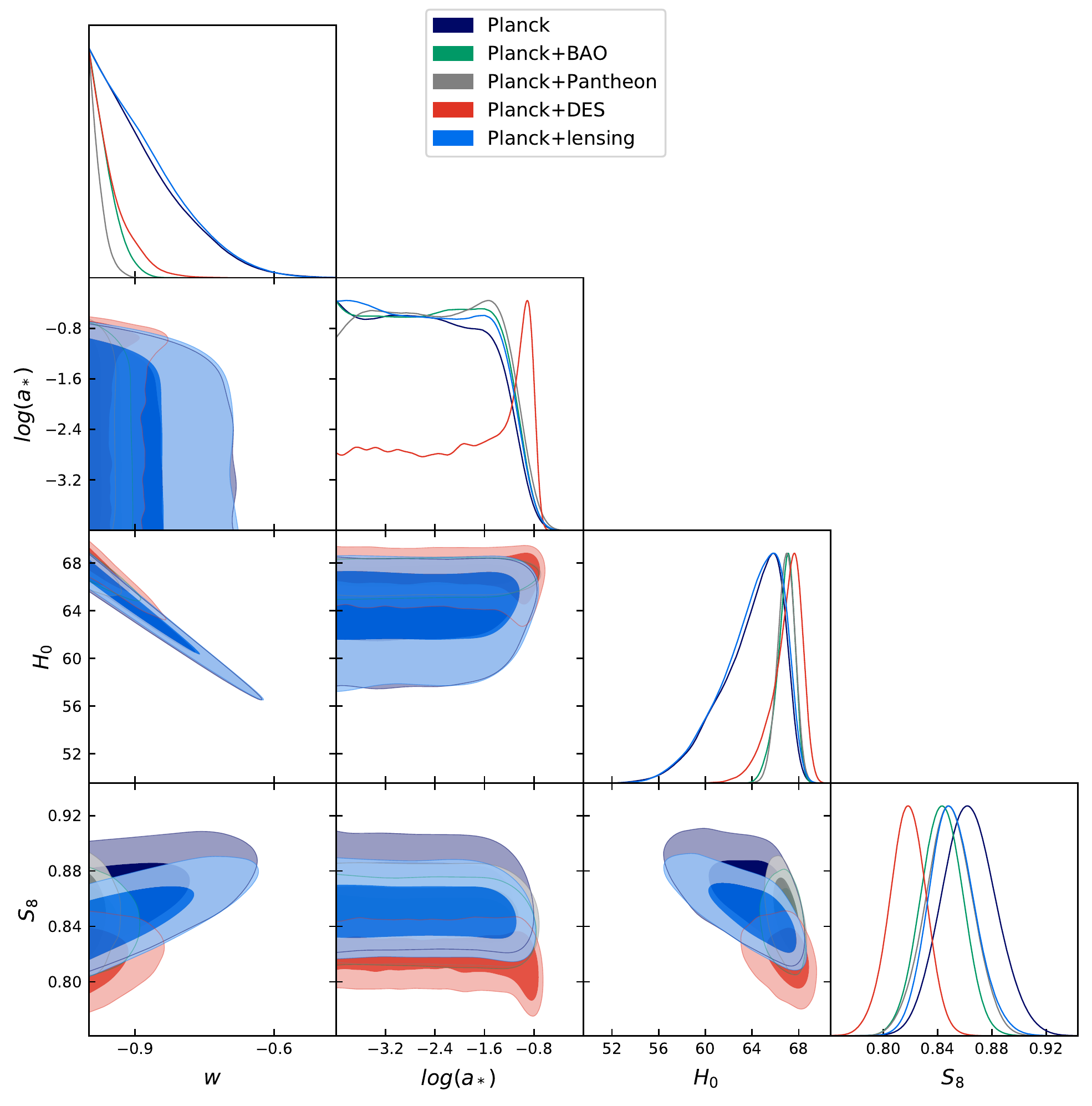}
\caption{2-D contours at 68$\%$~CL and 95$\%$~CL, and 1-D posterior distributions on some selected cosmological parameters, obtained by considering a dark energy equation of state free to vary.}
\label{Fig:w}
\end{figure}

\begin{figure}
\includegraphics[width=0.5\textwidth]{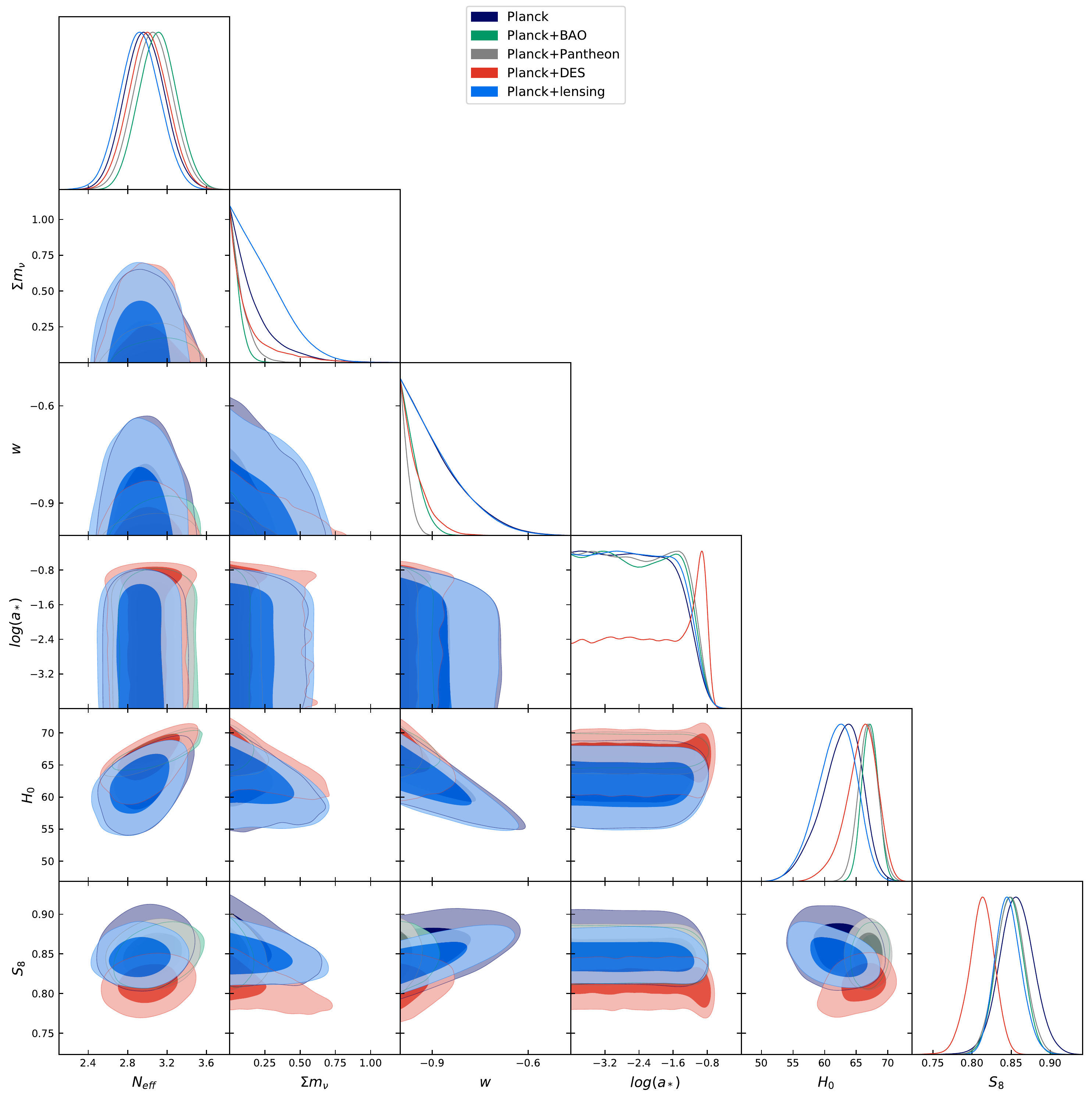}
\caption{2-D contours at 68$\%$~CL and 95$\%$~CL, and 1-D posterior distributions on some selected cosmological parameters, obtained by considering a dark energy equation of state and the neutrino parameters free to vary.}
\label{Fig:nu}
\end{figure}

We now present the results of our analyses combining the \textit{Planck} data with the cosmological probes considered in this work. We have considered four different analyses, in which we vary I) $a_*$ alone, see Table~\ref{a*}, II) $a_*$, $\Delta$, and $f_L$, see Table~\ref{fL}, III) $a_*$ and $w_{\phi 0}$, see Table~\ref{w}, and IV) $a_*$ and $w_{\phi 0}$ plus the total neutrino mass $\Sigma m_\nu$ and the number of relativistic degrees of freedom $N_{\rm eff}$, see Table~\ref{nu}. For each table, we show the $68 \% $ confidence level (CL)  limits on the cosmological parameters and we consider different combinations of the datasets used, with increased level of complexity. Moreover, we display the 2-D contours at 68$\%$~CL and 95$\%$~CL as well as the 1-D posterior distributions of some selected parameters in Figs.~\ref{Fig:a*}, ~\ref{Fig:delta}, ~\ref{Fig:w} and ~\ref{Fig:nu}.

%\squeezetable                                    
\begin{table*}
\begin{center}                              
\scriptsize
\begin{tabular}{cccccccccccccccc}       
\hline\hline                                                                                                                    
Parameters & \textit{Planck}   & \textit{Planck} & \textit{Planck}& \textit{Planck} & \textit{Planck} & \\ 
 &   & +BAO  & +Pantheon & +DES & + lensing \\ \hline
 
 $\Omega_b h^2$ & $    0.02224 \pm 0.00016$ &  $    0.02230\pm 0.00014$ & $    0.02226\pm 0.00015$ & $    0.02241\pm 0.00015$ & $    0.02225\pm 0.00015$ \\
 
$\Omega_c h^2$ & $    0.1199\pm 0.0014$  & $    0.1190\pm 0.0010$ & $    0.1195\pm 0.0013$ & $    0.1173\pm 0.0012 $& $    0.1194\pm 0.0014 $\\

$\tau$ & $    0.078\pm 0.017$ &  $    0.083\pm 0.016$ &  $    0.081\pm 0.017$ & $    0.075\pm 0.017$ & $    0.067\pm 0.013$\\

$n_s$ & $    0.9642\pm 0.0047$ & $    0.9668\pm 0.0040$ &   $    0.9653\pm 0.0045$ & $    0.9695\pm 0.0047$ &  $    0.9648\pm 0.0045$\\

${\rm{ln}}(10^{10} A_s)$ & $    3.092\pm 0.033$ &  $    3.100\pm 0.031$ & $    3.095\pm 0.032$ & $    3.079\pm 0.032$ & $    3.068\pm 0.025$\\

$\sigma_8$ & $    0.830\pm 0.013$ &  $    0.831\pm 0.013$ & $    0.830\pm 0.013$ & $    0.815^{+0.013}_{- 0.012}$ & $    0.8185\pm 0.0087$\\

$H_0 $[km/s/Mpc] & $   67.22\pm 0.64$&  $   67.64\pm 0.47$ & $   67.41\pm 0.60$ & $   68.37\pm 0.56$ & $   67.42\pm 0.62$\\

$S_8$ & $    0.852\pm 0.018$ &  $    0.845\pm 0.016$ & $  0.849\pm 0.017  $ & $    0.815\pm 0.014$ & $    0.836\pm 0.013$\\

$\log_{10}(a_*)$ & $   < -1.07$ &  $    <-1.09$ & $  <-1.08  $ & $    <-0.812$ & $    <-1.01$\\

\hline\hline                                                  
\end{tabular}                                                   \caption{Measurements at 68$\%$~CL errors on the cosmological parameters using different combinations of the cosmological datasets considered here, obtained by fixing $f_L = 1$ and $\Delta=0.5$. The upper limits of $\log_{10}(a_*)$ are instead at 95$\%$~CL.}
\label{a*}
\end{center}    
\end{table*}

The first thing that we can notice when looking at the Table~\ref{a*} is that the constraints on the cosmological parameters from \textit{Planck} alone are exactly the ones obtained considering a $\Lambda$CDM model~\cite{Ade:2015xua}. In fact, the analysis only leads to an upper limit for the quantity $\log_{10}(a_*)$ that is around $a_*<10^{-1}$ at 95$\%$~CL. This result can be understood by looking at the effect on the temperature and polarization power spectra obtained by varying $a_*$ and showed in Fig.~\ref{fig:a*}, that are almost indistinguishable for $a_* < 10^{-1}$, providing the same fit of the CMB data. Moreover, the new parameter $a_*$ is not correlated with the other cosmological parameters, and in particular with $H_0$ and $S_8 \equiv\sigma_8 \sqrt{\Omega_m/0.3}$, as we can see in Fig.~\ref{Fig:a*}. Therefore, the introduction of a quintessence model with an effective equation of state parametrized by Eq.~\ref{eq:weff} does not help in relieving the well known tensions between the \textit{Planck} measures in a $\Lambda$CDM scenario and SH0ES~\cite{Riess:2019cxk} on the Hubble constant, and \textit{Planck} and the cosmic shear data KiDS-450~\cite{Kuijken:2015vca,Hildebrandt:2016iqg,Conti:2016gav}, DES~\cite{Abbott:2017wau,Troxel:2017xyo} and CFHTLenS~\cite{Heymans:2012gg, Erben:2012zw,Joudaki:2016mvz} on $S_8$~\cite{DiValentino:2018gcu}.

The comparison between the different combination of datasets shows that the constraints on $a_*$ and on the parameters of the $\Lambda$CDM model are almost the same also when considering different observables. The most different bounds we have, slightly shifted with respect to the other cases but always in agreement with them, are those obtained from \textit{Planck} + DES. This is true for all the extended model considered in this work. 
%\squeezetable                                    
\begin{table*}
\begin{center}                              
\scriptsize
\begin{tabular}{cccccccccccccccc}       
\hline\hline                                                                                                                    
Parameters & \textit{Planck}   & \textit{Planck} & \textit{Planck}& \textit{Planck} & \textit{Planck} \\ 
 &   & +BAO  & +Pantheon & +DES & + lensing \\ 
 \hline
 
 $\Omega_b h^2$ & $    0.02223 \pm 0.00015$ &  $    0.02229\pm 0.00014$ & $    0.02226\pm 0.00015$ & $    0.02240\pm 0.00015$ & $    0.02225\pm 0.00016$\\
 
$\Omega_c h^2$ & $    0.1199\pm 0.0014$  & $    0.1190\pm 0.0011$ & $    0.1194\pm 0.0014$ & $    0.1172\pm 0.0013 $& $    0.1194\pm 0.0014 $ \\

$\tau$ & $    0.078\pm 0.017$ &  $    0.083\pm 0.017$ &  $    0.081\pm 0.017$ & $    0.075\pm 0.017$ & $    0.067\pm 0.014$\\

$n_s$ & $    0.9641\pm 0.0048$ & $    0.9662\pm 0.0041$ &   $    0.9654\pm 0.0046$ & $    0.9696\pm 0.0046$ &  $    0.9649\pm 0.0047$ \\

${\rm{ln}}(10^{10} A_s)$ & $    3.092\pm 0.033$ &  $    3.100\pm 0.033$ & $    3.096\pm 0.033$ & $    3.078\pm 0.032$ & $    3.068\pm 0.026$\\

$\sigma_8$ & $    0.830\pm 0.013$ &  $    0.831\pm 0.013$ & $    0.831\pm 0.013$ & $    0.816\pm 0.012$ & $    0.8188\pm 0.0091$\\

$H_0 $[km/s/Mpc] & $   67.20\pm 0.64$&  $   67.60\pm 0.47$ & $   67.42\pm 0.62$ & $   68.35\pm 0.59$ & $   67.41\pm 0.65$\\

$S_8$ & $    0.852\pm 0.017$ &  $    0.846\pm 0.016$ & $  0.849\pm 0.017  $ & $    0.816\pm 0.013$ & $    0.837\pm 0.013$\\

$f_L$ & unconstrained &  unconstrained & unconstrained & unconstrained & unconstrained\\

$\Delta$ & $  >0.250$ &  $  >0.250$ & $  >0.261$ & $  >0.238$ &     $  >0.254$\\

$\log_{10}(a_*)$ & $   -2.2 ^{+1.3}_{-0.8}$ &  $    2.2_{-    0.8}^{+    1.3}$ & $  2.3_{-    0.8}^{+    1.3}  $ & $    -2.06_{-  0.9}^{+    1.4}$ & $    -2.2_{-    0.8}^{+    1.3}$\\

\hline\hline                                                  
\end{tabular}                                                   \caption{Measurements at 68$\%$~CL errors on the cosmological parameters using different combinations of the cosmological datasets considered here. The lower limits of $\Delta$ are instead at 95$\%$~CL.}
\label{fL}
\end{center}
\end{table*}

Since in the baseline model we fix $f_L = 1$ and $\Delta=0.5$, we tested the robustness of our results by letting free to vary these two additional parameters of the model. We can see in Table~\ref{fL} the results obtained in this way. Also in this case we can notice that the constraints on the cosmological parameters from \textit{Planck} alone are exactly the ones obtained considering a $\Lambda$CDM model, and the previous case with $f_L$ and $\Delta$ fixed. However, while $f_L$ is completely unconstrained for all the combination of datasets considered in this work, we find now a constraint at 68$\%$~CL for $\log_{10}(a_*)$, thanks to its slight correlation with $\Delta$, as we can see in Fig.~\ref{Fig:delta}. Also in this case, these new parameters $f_L$ and $\Delta$ do not correlate with $H_0$ and $S_8$, as we can see in Fig.~\ref{Fig:delta}, and therefore do not alleviate the tensions between \textit{Planck} and the other cosmological probes.

%\squeezetable                                    
\begin{table*}
\begin{center}                              
\scriptsize
\begin{tabular}{cccccccccccccccc}       
\hline\hline                                                                                                                    
Parameters & \textit{Planck}   & \textit{Planck} & \textit{Planck}& \textit{Planck} & \textit{Planck} & \\ 
 &   & +BAO  & +Pantheon & +DES & + lensing \\ \hline
 
 $\Omega_b h^2$ & $    0.02221 \pm 0.00016$ &  $    0.02234\pm 0.00014$ & $    0.02228\pm 0.00014$ & $    0.02241\pm 0.00015$ & $    0.02223\pm 0.00016$ \\
 
$\Omega_c h^2$ & $    0.1201\pm 0.0015$  & $    0.1184\pm 0.0011$ & $    0.1193\pm 0.0014$ & $    0.1173\pm 0.0012 $& $    0.1197\pm 0.0014 $\\

$\tau$ & $    0.080\pm 0.017$ &  $    0.086\pm 0.017$ &  $    0.082\pm 0.017$ & $    0.075\pm 0.017$ & $    0.071\pm 0.014$\\

$n_s$ & $    0.9639\pm 0.0048$ & $    0.9680\pm 0.0042$ &   $    0.9656\pm 0.0048$ & $    0.9693\pm 0.0048$ &  $    0.9644\pm 0.0047$\\

${\rm{ln}}(10^{10} A_s)$ & $    3.096\pm 0.033$ &  $    3.104\pm 0.032$ & $    3.097\pm 0.033$ & $    3.079\pm 0.033$ & $    3.077\pm 0.026$\\

$\sigma_8$ & $    0.798^{+0.034}_{- 0.019}$ &  $    0.821\pm 0.015$ & $    0.825\pm 0.014$ & $    0.818^{+0.020}_{- 0.014}$ & $    0.789^{+0.030}_{-0.016}$\\

$H_0 $[km/s/Mpc] & $   63.9^{+3.3}_{-1.5}$&  $   66.86^{+0.89}_{-0.59}$ & $   66.92^{+0.79}_{-0.69}$ & $   67.0^{+1.6}_{-0.8}$ & $   63.9^{+3.4}_{-1.5}$\\

$S_8$ & $    0.863\pm 0.020$ &  $    0.843\pm 0.015$ & $  0.848\pm 0.017  $ & $    0.818\pm 0.014$ & $    0.851^{+0.016}_{-0.018}$\\

$w_{\phi 0}$ & $   <-0.709$ &  $    <-0.912$ & $  <-0.945  $ & $    <-0.872$ & $    <-0.707$\\

$\log_{10}(a_*)$ & $   <-1.14$ &  $    <-1.10$ & $  <-1.06  $ & $    -2.1_{-  1.7}^{+    1.4}$ & $    <-1.11$\\

\hline\hline                                                  
\end{tabular}                                                   \caption{Measurements at 68$\%$~CL errors on the cosmological parameters using different combinations of the cosmological datasets considered here, considering a free dark energy equation of state. The upper limits of $w$ and $\log_{10}(a_*)$ are instead at 95$\%$~CL.}
\label{w}
\end{center} 
\end{table*}

In order to test the stability of our results, we tested the addition of a quintessence dark energy equation of state free to vary. In this case we fixed again $f_L = 1$ and $\Delta=0.5$, because they do not add any additional information, and these values are perfectly consistent within one standard deviation with our findings in the previous case. We can see in Table~\ref{w} the results obtained for this extended scenario. In this case we have just an upper limits for both $w_{\phi 0}$ and $\log_{10}(a_*)$. However, because of the strong anti-correlation between $w_{\phi 0}$ and $H_0$, we find a shift of the Hubble constant towards lower values, increasing the tension with \textit{Planck} and the local measurements from SH0ES~\cite{Riess:2019cxk}. 
Moreover, we have an indication at one standard deviation for $\log_{10}(a_*)$ for the \textit{Planck}+DES combination.

%\squeezetable                                    
\begin{table*}
\begin{center}                              
\scriptsize
\begin{tabular}{cccccccccccccccc}       
\hline\hline                                                                                                                    
Parameters & Planck   & Planck & Planck& Planck & Planck & \\ 
 &   & +BAO  & +Pantheon & +DES & + lensing \\ \hline
 
 $\Omega_b h^2$ & $    0.02211 \pm 0.00025$ &  $    0.02237\pm 0.00020$ & $    0.02229\pm 0.00023$ & $    0.02234\pm 0.00024$ & $    0.02205\pm 0.00026$ \\
 
$\Omega_c h^2$ & $    0.1193\pm 0.0031$  & $    0.1195\pm 0.0031$ & $    0.1194\pm 0.0031$ & $    0.1170\pm 0.0029 $& $    0.1185\pm 0.0031 $\\

$\tau$ & $    0.081\pm 0.018$ &  $    0.087\pm 0.017$ &  $    0.083\pm 0.018$ & $    0.075\pm 0.017$ & $    0.077\pm 0.017$\\

$n_s$ & $    0.9601\pm 0.0098$ & $    0.9699\pm 0.0082$ &   $    0.9662\pm 0.0092$ & $    0.9671\pm 0.0094$ &  $    0.9579\pm 0.0098$\\

${\rm{ln}}(10^{10} A_s)$ & $    3.094\pm 0.038$ &  $    3.108\pm 0.036$ & $    3.100\pm 0.038$ & $    3.078\pm 0.038$ & $    3.085\pm 0.036$\\

$\sigma_8$ & $    0.776^{+0.049}_{-0.029}$ &  $    0.827\pm 0.019$ & $    0.824^{+0.022}_{-0.020}$ & $    0.720^{+0.045}_{- 0.021}$ & $    0.758^{+0.042}_{-0.029}$\\

$H_0$ [km/s/Mpc]& $   62.6^{+3.7}_{-2.5}$&  $   67.2\pm 1.3$ & $   66.9\pm 1.5$ & $   65.8^{+2.8}_{-2.0}$ & $   61.8^{+3.5}_{-2.7}$\\

$S_8$ & $    0.857\pm 0.022$ &  $    0.848\pm 0.017$ & $  0.849\pm 0.018  $ & $    0.812^{+0.018}_{-0.015}$ & $    0.846^{+0.015}_{-0.018}$\\

$w_{\phi 0}$ & $   <-0.703$ &  $    <-0.903$ & $  <-0.946  $ & $    <-0.867$ & $    <-0.712$\\

$\Sigma m_{\nu} [{\rm eV}]$ & $   <0.516$ &  $    <0.136$ & $  <0.215  $ & $    <0.549$ & $    <0.564$\\

$N_{\rm eff}$ & $   2.97\pm0.20$ &  $    3.11\pm0.19$ & $  3.06\pm0.20  $ & $    3.01\pm0.20$ & $    2.92\pm0.20$\\

$\log_{10}(a_*)$ & $   <-1.16$ &  $    <-1.11$ & $  <-1.08  $ & $    -2.2_{-  1.6}^{+    1.4}$ & $    <-1.14$\\

\hline\hline                                                  
\end{tabular}
\caption{Measurements at 68$\%$~CL errors on the cosmological parameters using different combinations of the cosmological datasets considered here, considering free parameters in the neutrino sector. The upper limits of $w_{\phi 0}$, $\Sigma m_{\nu}$ and $\log_{10}(a_*)$ are instead at 95$\%$~CL.}
\label{nu}
 \end{center} 
 \end{table*}

As a final step, we investigated the effect of our model on the parameters of the neutrino sector, in particular the total neutrino mass $\Sigma m_\nu$ and the number of relativistic degrees of freedom $N_{\rm eff}$ in Table~\ref{nu}. Again, we only find an upper limit for the quantities $w_{\phi 0}$, $\Sigma m_\nu$ and $\log_{10}(a_*)$, while $N_{\rm eff}$ is perfectly in agreement with its standard value $3.046$~\cite{Mangano:2005cc, deSalas:2016ztq}. Due to the anti-correlation between $w_{\phi 0}$ and $H_0$ and $\Sigma m_\nu$ and $H_0$, we find a further shift of the Hubble constant towards lower values, increasing the tension with \textit{Planck} and the local measurements~\cite{Riess:2019cxk}. 
Also in this extended scenario, we notice an indication at one standard deviation for $\log_{10}(a_*)$ for the \textit{Planck}+DES combination.

\section{Conclusions}
\label{sec:Conclusions}

In this work we have investigated an extension of the $\Lambda$CDM model that includes an additional quintessence field $\phi$ whose equation of state changes during the late-time evolution of the Universe. We have modelled the equation of state of the quintessence field to behave as a matter field around recombination, $w_\phi(a_{\rm recomb}) = 0$, transitioning to a generic equation of state $w_{\phi 0}$ at a later time. In the first part of this work, we have set $w_{\phi 0} = -1$, so that the quintessence field behaves as a cosmological constant today, while in the second part, where we considered extended models, we have let it free to vary. We have modified the publicly-available code \texttt{CAMB} to perform a MCMC analysis and assess the parameter space for the model, which extends the parameter space of the $\Lambda$CDM model to include the fraction $f_L$ of dark energy today that is in the quintessence field, the scale factor at which the transition occurs $a_*$ and the duration of the transition $\Delta$.

We obtained results for the datasets in two different setups, as summarized in Table~\ref{a*} in which we scan over the parameter $a_*$ while fixing $f_L = 1$ and $\Delta = 0.5$, and in Table~\ref{fL} where we include all three extra parameters of the model in the minimization. We used flat priors on the cosmological parameters as illustrated in Table~\ref{priors}. As the results in the tables show, our model does not help in alleviating the tension that is present in the $H_0$ and $S_8$ parameters. In particular, we find that data constraints the transition in such dark energy models to happen before $\log_{10}(a_*)\lesssim -1$ and are completely ineffective at constraining the parameter $f_L$.

Our results further suggest that early time changes in the cosmological history are more likely to explain the present tensions between different cosmological datasets\footnote{During the conclusion of this work the preprint \cite{Agrawal:2019lmo} appeared on the arXiv. Although the model has many similarities with the one discussed here, a quintessence field which transitions from dark matter to dark energy, the authors of \cite{Agrawal:2019lmo} claim a more positive result. Note, however, that, as mentioned in section \ref{sec:Method}, we have not used in this work datasets with more than $3\sigma$ tensions between themselves.}.

\appendix

\section*{Acknowledgements}
We thank Prateek Agrawal, Rahul Biswas, Fabrizio Rompineve, Ben Safdi, Sunny Vagnozzi and Licia Verde for useful suggestions that improved the paper. EDV acknowledges support from the European Research Council in the form of a Consolidator Grant with number 681431. LV acknowledges support by the Vetenskapsr\r{a}det (Swedish Research Council) through contract No.~638-2013-8993 and the Oskar Klein Centre for Cosmoparticle Physics. UD acknowledges support by the Vetenskapsr\r{a}det (Swedish Research Council) under contract 2015-04814. LV thanks the kind hospitality of the INFN Laboratori Nazionali di Frascati, the Leinweber Center for Theoretical Physics, and the University of Michigan, where part of this work was carried out.

\section{Tunneling probability} \label{sec:tunneling}

Here, we consider the quintessence model presented in Sec.~\ref{sec:realization}. The field is likely to get trapped in the false minima if the parameter $\kappa \equiv \Lambda^2/mf \gg 1$~\cite{Jaeckel:2016qjp}, which is a necessary condition for the field to transition to a cosmological constant behaviour. However, in order for the quintessence model presented to be a good candidate for the dark energy observed, we should ensure that the probability to tunnel to the next minima is small, otherwise, bubbles of the new vacua would start to form and collide leading to very different consequences. In this section we will study these probabilities in more detail following similar methods to those used in Ref.~\cite{Hebecker:2016vbl}.

In general, there are several channels for the scalar field to tunnel to the next minima. If the field is held in the metastable minima thermally, like for example in Ref.~\cite{Baratella:2018pxi}, thermal fluctuations need to be taken into account. The field can also tunnel through the formation of a Coleman de-Luccia bubble~\cite{Coleman:1980aw} but that has been shown to be unlikely in this model~\cite{Kobayashi:2018nzh}. The third possibility, is that the sole presence of density perturbations, generated for example from inflation, can make the field jump over the barrier. We will focus on this last possibility here.

Classically, the field will stop in a minima after it lost enough kinetic energy in half period to not overcome the barrier. However, quantum fluctuations can still allow it to jump to the other side. Therefore, we should ensure that the energy lost in half period, $\Delta \rho$, is larger than the energy fluctuations of the field, i.e.~\cite{Hebecker:2016vbl}
\begin{eqnarray}
\frac{\delta \rho_\phi (k,t)}{\Delta \rho} \ll 1 \, . \label{inflationary perturbations condition}
\end{eqnarray} 

In the case of adiabatic density perturbations of inflationary origin, because they are conserved on superhorizon scales, they can be fixed at horizon crossing (hc) to $\delta \rho_\phi (t_\text{hc})/\rho_\phi = \sqrt{P_\zeta}/3$ where $P_\zeta =2.2 \times 10^{-9}$ is the amplitude of the primordial scalar power spectrum~\cite{Akrami:2018odb}. Once inside the horizon, fluctuations grow logarithmically during radiation domination and linearly during matter domination, leading to
\begin{eqnarray}
    \frac{\delta \rho_\phi (k,t)}{\rho_\phi} \sim \frac{\sqrt{P_\zeta}}{3} \frac{a(t)}{a_\text{eq}} \left[1+ \ln \left(\frac{k}{a_\text{hc} H_\text{eq}} \right) \right] \, 
\end{eqnarray}
where we assumed that the mode of interest $k \simeq m$ has entered the horizon during radiation domination.

Regarding $\Delta \rho$ it can be estimated by noting that $\rho \propto t^{-2}$. Then,
\begin{eqnarray}
\frac{\Delta \rho_\phi}{\rho_\phi} \sim 2 \frac{\Delta t}{t} \sim \frac{3H}{M}
\end{eqnarray}
where $M$ is the curvature around the minima which sets the period of the oscillation.
Now, after using that $H_*= H_\text{eq} (a_\text{eq}/a_*)^{3/2}$ during matter domination and neglecting the logarithmic term we find that Eq.~\ref{inflationary perturbations condition} requires
\begin{eqnarray}
    \frac{H_*}{H_\text{eq}} \gtrsim \left(\frac{P_\zeta}{9}\right)^{3/10} \left(\frac{M}{H_\text{eq}}\right)^{3/5} 
    \simeq  7\times 10^{-4} \left(\frac{M}{H_\text{eq}}\right)^{3/5} \, . \label{inflationary perturbations condition 2}
\end{eqnarray}
This provides a non-trivial constraint on the model. For example, for $\kappa=10$, requiring $m \gtrsim H_\text{eq}$ gives
\begin{eqnarray}
\frac{H_*}{H_\text{eq}} \gtrsim  3\times 10^{-3}  \qquad \Rightarrow \qquad a_* < 0.07 \,.  
\end{eqnarray}
Note, however, that if the density perturbations were not adiabatic then, given the present constraint on isocurvature perturbations \cite{Akrami:2018odb}, the probability to jump would be smaller. So we take Eq.~\ref{inflationary perturbations condition 2} as a conservative bound.

\section*{References}

\bibliographystyle{elsarticle-num}
\bibliography{H0Tension.bib}

\end{document}